%% file: main.tex
\documentclass[12pt]{iopart}
\usepackage{xcolor, soul}

\sethlcolor{green}
\usepackage{hyperref}
\usepackage{booktabs}
\usepackage{float}
\usepackage[flushleft]{threeparttable}
\usepackage{iopams}

\usepackage[nolist]{acronym}
\usepackage{graphicx} 
\usepackage{hhline}
\usepackage[T1]{fontenc} 
\usepackage{makecell}
\usepackage{multirow}
\usepackage{pdflscape}
\usepackage{rotating}
\usepackage{subcaption}
\usepackage{subfiles}
\usepackage{textcomp}
\usepackage{tikz}
\usetikzlibrary{3d, patterns, patterns.meta, shapes.geometric, arrows.meta, positioning}
\usepackage{todonotes}

\newcommand{\refDose}{d^{\textsuperscript{ref}}}
\newcommand{\R}{\mathbb{R}} 
\newcommand{\numObj}{N} 

\definecolor{ITWM}{cmyk}{1.0, 0, 0.65, 0} 
\definecolor{MGH}{rgb}{1.0, 0.65, 1}

\begin{acronym}
    \acro{CTV}{clinical target volume}
    \acro{DVH}{dose-volume histogram}
    \acro{EUD}{equivalent uniform dose}
    \acro{MCO}{multi-criteria optimization}
    \acro{NSCLC}{non small cell lung cancer}
    \acro{NTCP}{normal tissue complication probability}
    \acro{OAR}[OAR]{organ at risk}
    \acroplural{OAR}[OARs]{organs at risk}
    \acro{PTV}{planning target volume}
    \acro{RP}{radiation pneumonitis}
    \acro{RT}{radiation therapy}
    \acro{TCP}{tumor control probability}
\end{acronym}

\begin{document}
\title{Bi-level Multi-criteria Optimization for Risk-informed Radiotherapy}
\author{$^{1,*}$Mara Schubert, $^1$Katrin Teichert, $^2$Zhongxing Liao, $^3$Thomas Bortfeld, $^3$Ali Ajdari}
$^1$ Fraunhofer Institute for Industrial Mathematics (ITWM), Fraunhofer-Platz 1, 67663 Kaiserslautern, Germany\\
$^2$ University of Texas‘ MD Anderson Cancer Center, Houston, USA\\
$^3$ Department of Radiation Oncology, Massachusetts General Hospital \& Harvard Medical School, Boston, USA\\
\ead{mara.schubert@itwm.fraunhofer.de}
\vspace{10pt}
\begin{indented}
\item[]\today
\end{indented}

\begin{abstract}
\subfile{Abstract}

\end{abstract}

	
	
	
\section{Introduction}
\subfile{Introduction}
\section{Materials and Methods}

\subfile{MaterialMethods}

\section{Results}
\subfile{Results}

\section{Discussion}
\subfile{Discussion}

\section{Conclusion}
\subfile{Conclusion}

\section{Appendix}
\subfile{Appendix}

\section*{Competing interests}

\bibliographystyle{apalike} 
\bibliography{references}

\end{document}

%% file: Abstract.tex
\textbf{Purpose:} In \ac{RT} treatment planning, \ac{MCO} allows the physician to find the best plan efficiently. \Ac{MCO} is conventionally solved for a set of generic (population-wide) dosimetric criteria, ignoring patient-specific biological risk factors and compromising clinical outcomes in high-risk groups. We propose a one-shot method - risk-guided \ac{MCO} - for direct integration of biological risk factors within conventional \ac{MCO}, enabling interactive plan navigation between dosimetric and biological endpoints.

\textbf{Methods:} A cohort of \ac{NSCLC} patients receiving proton/photon RT was retrospectively analyzed. The clinical endpoint was the risk of symptomatic (grade 2+) \ac{RP}, modeled using bootstrapped stepwise logistics regression (with interactions) accounting for baseline lung function, smoking history, and conventional dosimetric factors. We defined a special order relation to fuse the conventional \ac{MCO} sandwiching algorithm with bi-level optimization, restricting the (infinite) Pareto set to plans with substantial gain in the secondary risk objective for acceptable loss in primary (clinical) objectives. Thus, risk-guided \ac{MCO} computes risk-optimized counterparts of clinical plans in a single run (rather than a sequential/lexicographic approach) within user-defined trade-offs. Performance was assessed in terms of clinical objectives and predicted \ac{RP} risk.

\textbf{Results:} 
Across 19 patients, the risk-guided plans yielded an 8.0\% mean reduction in total lung V20 and 9.5\% reduction in right lung V5, translating into an average \ac{RP} risk reduction of 7.7\% (range=0.3\%-20.1\%), with only small changes in target coverage (mean -1.2 D98[\%] for CTV) and modest increase in heart dose (mean +1.74 Gy).



\textbf{Conclusions:} This study presents the first proof-of-concept for integrating biological risk models directly within multi-criteria \ac{RT} planning, enabling an interactive balance between established population-wide dose protocols and individualized outcome prediction. Our results demonstrate that the risk-informed
\ac{MCO} can reduce the risk of RP while maintaining target coverage.

%% file: Introduction.tex




Cancer \ac{RT} is fundamentally a multi-objective decision making process, aiming on the one hand to control cancer progression, and on the other to limit treatment-induced side effects. These two goals are often in conflict with one-another, as increasing the radiation dose would often lead to higher chances of cancer control and simultaneously increased rate of radiation-induced toxicity (RIT). This is further complicated by the multitudes of clinically important side-effects that need to be considered as part of \ac{RT} treatment design. In \ac{NSCLC} treatment, for example, the risk of multiple side-effects, primarily \ac{RP}, esophagitis, and cardiovascular disease (CVD), demand careful clinical attention and should be balanced against the desire to control tumor growth.  Therefore, to design an effective \ac{RT} plan, one has to carefully navigate this high-dimensional space, weighing the potential benefit and harm of a higher radiation dose both in terms of \ac{TCP} as well as the risk of \ac{NTCP}.

\sloppy \Ac{MCO} is an established approach for RT treatment planning and has been implemented in commercial treatment planning systems (RayStation\texttrademark\ by RaySearch Laboratories \cite{RayMCO17}, Eclipse\texttrademark\ by Varian \cite{VarMCO19}). When using \ac{MCO}, the planner must define objectives and constraints that reflect the required target coverage and sparing of \acp{OAR}. Subsequently, a plan data base of representative points is automatically generated, capturing the different feasible trade-offs between the objectives while ensuring that all clinically-important constraints are satisfied. In \ac{MCO} parlance, each representative plan in the data base is \emph{Pareto optimal}, which means that no objective can be improved without worsening at least one other objective.

In conventional \ac{MCO} the objectives of both, target and \acp{OAR}, are represented using physical dose-based constraints which are based on clinical guidelines that stem from historic data and population-wide averages. However, such estimates are, by definition, tailored to hypothetical ``average'' patients and could prove to be poor surrogates for the individual patient's outcome. The response of patients, both in terms of tumor control and toxicity profile, is a complex process that is affected by several factors beyond radiation dose alone, including, among others, patients' pathophysiology, genetic and epigenetic makeup, and socioeconomic factors \cite{Li.2024}. In recent years, many studies have looked beyond dose-only factors to estimate the patient-specific response after radiotherapy. They include attempts to synthesize, alongside conventional dosimetrics, additional features from radiological images (i.e. radiomics) (\cite{Xiao.2025}, \cite{Ren.2023}, \cite{Peeken.2017}), molecular assays (\cite{Pan.2023}, \cite{Bera.2022}, \cite{Velu.2024}, \cite{Rabasco.2022}) and other clinicopathological features. Advancements in the field of artificial intelligence (AI) and machine learning have also contributed significantly to an ever-expanding landscape of biological risk/outcome models (\cite{Wei.2019}, \cite{Deist.2018}, \cite{Rydzewski.2023}, \cite{Yuan.2025}). 

Despite the promise of these more personalized outcome models, the field has largely lagged behind on the integration of those models within clinical decision making process. Most clinical applications still rely on broad risk-stratification frameworks, such as de-escalation strategies for HPV-positive head-and-neck cancer \cite{Petkar.2025}, rather than embedding individualized biological predictions directly into dose optimization. Early efforts toward ``biologically informed plannin'' were largely restricted to simplified radiobiological models—including classical Lyman–Kutcher–Burman NTCP formulationss—or handcrafted surrogates such as mean-lung-dose thresholds or parotid mean-dose limits (\cite{Wang.25}, \cite{Raturi.2021}, \cite{Kierkels.2016}). While these population-derived models have shaped practice, they do not leverage the richer structure of patient-specific data now available from radiomics, genomics, clinical biomarkers, and other multimodal predictors. Conceptual work on dose painting and theragnostic imaging further articulated how functional and molecular imaging could be used to prescribe heterogeneous dose distributions (\cite{Ureba.2022}, \cite{Toma-Dasu.2013}), but in practice these approaches still largely operate at the level of population-derived constraints rather than fully individualized biological risk. More recent efforts have moved toward explicitly risk-informed optimization, in which empirical or machine-learning models are coupled more tightly to planning. Ajdari et al. \cite{Ajdari_2022} demonstrated mid-course FDG-PET–based adaptation for NSCLC, where a Bayesian network and patient-specific lung radiosensitivity factor personalize an NTCP model and enable NTCP-guided plan adaptation at mid-treatment. Building on this, Maragno et al. \cite{Maragno_2024} embedded entire machine-learning toxicity models—decision trees, ensembles, and neural networks—within an ``optimization with constraint learning'' framework to directly constrain patient-specific pneumonitis risk during lung plan optimization. Despite these promising efforts, studies on integrating biological outcome models within RT planning remain predominantly proof-of-concept, leaving a large gap for practical, clinically deployable framework that integrates individualized biological risk models into the optimization workflow.

Specifically, two critical gaps remain in translating these risk-informed strategies into clinically viable decision tools. First, the majority of the current risk models in the literature have not undergone rigorous clinical testing and, as such, might lack the necessary clinical trust for them to be used to guide important clinical decisions. Second, improving the likelihood of a predicted outcome often requires accepting trade-offs in other, more established dose-based objectives. However, a risk-guided planning process in which risk-based criteria are just as likely to drive the outcome as more established dosimetric criteria inevitably run the risk of deviating too much from acceptable clinical protocols, rendering the resulting plan unsuitable for clinical purposes. To address these two gaps, we propose a novel solution wherein the risk-based criteria would be treated as \emph{secondary priority}, subordinate to established dose objectives. 

Our proposed method occupies a middle ground in the spectrum of clinical decision-making. On the one hand, conventional \ac{MCO} offers substantial freedom to explore patient-specific compromises from a wide range of options, allowing flexibility without the need for extensive prior definition of clinical goals. However, as the number of objectives increases, the interactive decision-making process can become highly complex for the planner.
On the other hand, the lexicographic approach of ICycle \cite{BreStoVoe12} requires the establishment of a prioritized ranked list of objectives and constraints—a so-called wish-list—as a means of assessing plan quality. Although ICycle is efficient and straightforward to use, it requires meticulous definition of the wish-list, which can hinder its adaptability to patient-specific factors such as risk.
Our proposed method incorporates prioritized objectives and automated evaluation of trade-offs into conventional \ac{MCO}. Specifically, our approach automatically finds risk-optimized plans within a user-defined neighborhood of the clinically-optimized Pareto front. It minimizes risk as much as possible while ensuring that any degradation in the plan with respect to dose objectives is only considered if the risk-improvement is high enough. Rooted in bi-level optimization, this strategy provides the flexibility to balance the trade-off between population-wide dose objectives to ensure clinical acceptability and model-based risk objectives to account for personalized biological response. 

We build our methodology based on a \ac{NSCLC} cohort (\cite{Ajdari_2022}, \cite{Liao.2018})  with the goal of balancing the risk of \ac{RP} against conventional clinical criteria to derive risk-informed, personalized treatment plans. The methodology directly builds on our prior work \cite{Schubert.2025} wherein we demonstrated that a special class of bi-level optimization problems can be reformulated and solved as a single \ac{MCO} problem by altering the order relation that defines when a solution is considered optimal. As such, the reformulated bi-level optimization approach enables the simultaneous consideration of objectives with different priorities. We further extend our prior work on risk-guided optimization of \ac{NSCLC} \ac{RT} plans \cite{Maragno_2024} to balance data-driven risk models against traditional dosimetric factors, thereby improving clinical acceptability. 

%% file: MaterialMethods.tex
\subsection{The risk-agnostic model}
\label{sec:risk_agnostic_model}
As a baseline for the assessment of risk-optimized plans, we calculate plans optimized solely for dose objectives. These plans are henceforth referred to as risk-agnostic, since no \emph{direct} attention is paid to the individualized risk prediction in the optimization. We use a multi-criteria setting that allows the exploration of trade-off between different dose objectives $F_i: \R^n \rightarrow \R$, $i= 1,..,\numObj$. As such, the risk-agnostic \ac{MCO} model is 
\begin{equation}
\eqalign{
    \min_{x \in \R^m} \quad&F(d)\cr
    \quad \mathrm{s.t.} \quad & G(d) \leq 0, \cr
    &d=Dx, \\ 
    &x\geq 0}
 \label{eq:risk-agnostic-MCO}
\end{equation}
where $x$ is the plan, $d$ is a vector that describes the dose per voxel, $F(d)$ and $G(d)$ are vector valued functions, and $D \in \mathbb{R}^{n \times m}$ is the dose-influence matrix. The solutions of \eref{eq:risk-agnostic-MCO} include all feasible points where no other feasible point is at least equally good in all and better in one of the objectives $F_i(d)$.

\subsubsection{Dose-based objectives and constraints}

All objectives and constraints were modeled using one of the following dose metrics, each evaluating the dose distribution $d = d(x)$ in a given volume $V$, with $p$ denoting the norm exponent, and $\refDose$ denoting a threshold dose value.

\begin{itemize}
\item The \emph{equivalent uniform dose} (EUD) penalizes any dose in the volume:
\begin{equation}
    \mathrm{EUD}(d) = \frac{1}{\left| V \right|} \sum_{v \in V} d_v^p.
    \nonumber
\end{equation}
\item \emph{Overdose} (OD) and \emph{underdose} (UD) penalize any voxel dose above or below a certain threshold, respectively:
\begin{equation}
	\mathrm{UD}(d) = \frac{1}{\left| V \right|}\sum_{v \in V} \left( \max\left\lbrace 0, \refDose - d_v \right\rbrace\right)^p,
    \nonumber
\end{equation} 
and 
\begin{equation}
	\mathrm{OD}(d) = \frac{1}{\left| V \right|}\sum_{v \in V} \left( \max\left\lbrace 0, d_v - \refDose  \right\rbrace \right)^p.
    \nonumber
\end{equation}
\item The \emph{volume percentile} V$\left[\refDose\right]$ denotes the percentage of volume $V$ receiving dose above the threshold value $\refDose$: 
\begin{equation}
\mathrm{V}\left[\refDose\right] = \frac{100\%}{\left| V \right|}\sum_{v \in V} H(d_v-d^{ref})
\nonumber
\end{equation}
with $H$ denoting the heaviside step function. For numerical optimization, we employ the logistic approximation:
\begin{equation}
    \mathrm{V}\left[\refDose\right]= \frac{100\%}{\left| V \right|} \sum_{v \in V} \frac{1}{1+ e^{-2k(d_v-\refDose)}}
    \nonumber
\end{equation}
with $k=10$.
\end{itemize}

\subsubsection{Generating the Pareto optimal plans using conventional MCO}
\label{sec:POplans}
To facilitate the calculation of a single Pareto optimal treatment plan using a gradient-based numerical solver, the different competing planning objectives defined by the planner must be combined into a real-valued optimizable function $s$. This step is called the scalarization of the \ac{MCO} problem. A wide range of scalarization methods are described in the literature, with popular methods being norm- or direction-based approaches, weighted sum scalarization, and the epsilon constraint method (see for example \cite{Ehr05}, \cite{Eic09}). We employ the weighted sum method, where the objectives are individually weighted and summarized. 
The weighted sum scalarization for problem (\ref{eq:risk-agnostic-MCO}) is 
\begin{equation}
s(d) = w^T F(d)
\nonumber
\label{eq:weighted_sum_scal}
\end{equation}
with 
$w_i > 0, i=1,..,\numObj$. A plan that minimizes $s(d)$ while satisfying the constraints is Pareto optimal (see \cite{Ehr05}). 

For the generation of the plan database of Pareto optimal plans, we employed a polyhedral Sandwiching algorithm (\cite{LamKue25}, \cite{Ser12}, \cite{BokFor12}). This Sandwiching algorithm iteratively determines the weights for the next weighted sum problem to be solved based on the plans that have already been calculated (a pseudocode formulation can be found in \cite{BokFor12} Algorithm 3.1.). The Sandwiching algorithm guaranties that the generated plan database is a good representation of the Pareto front.



\subsection{Integrating the risk model}
\label{sec:integrating_the_risk_model}


\subsubsection{Risk model as secondary objective}
\label{subsec:secondary_objective}
A risk model evaluates the dose $d$ and assigns it a likelihood of risk realization, with the mapping being modified by patient individual characteristics. We call this patient-specific mapping from dose to risk the \emph{risk objective} and denote it by $r(d)$.
For each patient, we obtained the risk-guided front by adding the risk objective $r(d)$ as secondary priority after the dose objectives. For a fixed solution $x^*$ of \eref{eq:risk-agnostic-MCO} with dose $d^* = Dx^*$, this can be understood as the re-optimization 
\begin{equation}
\eqalign{
\min_{x\in \R^m}\quad & r(d)\cr
\quad \mathrm{s.t.}\quad & F(d)\le F(d^*)\cr
& G(d)\le 0\cr
& d = D x\cr
& x \ge 0}
\label{eq:reopt}
\end{equation}
Since we are not only interested in the re-optimization of a single solution but all solutions from \eref{eq:risk-agnostic-MCO} simultaneously, 
we apply the algorithm detailed in \cite{Schubert.2025}.
Therein, the algorithmic idea is to add the risk objective to the conventional \ac{MCO} problem, but to also modify the \emph{domination cone} of the \ac{MCO} problem to reflect the objective prioritization. 

In MCO, the domination cone $C$ can be used to describe whether a point $y$ is better than another point $z$. Namely, this holds whenever $y \in z - C$. Applying the algorithmic framework from \cite{Schubert.2025}, we make use of this description and define a particular domination cone that maintains strict prioritization between dose objectives (primary) and risk objective (secondary). From the cone that defines strict prioritization, we can derive an approximation cone $C_A$ that loosens the prioritization depending on a parameter $\epsilon$ (see Fig.\ref{fig:cones}), with strict prioritization achieved asymptotically for $\epsilon \rightarrow 0$ .  The modified \ac{MCO} problem reads as:
\begin{equation}
    \eqalign{
        C_A -\min_{x \in R^m} \quad &F(d), r(d) \cr
        \quad \quad \quad \mathrm{s.t.} \quad &G(d)\leq 0, \cr
        & d= Dx, \\ &x \geq 0.}
    \label{eq:risk-guided}
\end{equation}
For our case of a two dimensional dose objective and a one dimensional risk objective, the approximation cone is:
\begin{equation*}
		C_A= \{Q\lambda \mid \lambda \geq 0\},    
		\label{eq:CA}
	\end{equation*}
	with
	\begin{equation*}
		Q = \left( \begin{array}{ccccc}
		1& 0 &0& \epsilon &0\\
        0 &1 &0& 0 &\epsilon \\
        0 &0 &1 &-1 &-1     
		\end{array} \right).
		\label{eq:Q}
	\end{equation*}
The set of all points that are considered ideal compromises for this modified MCO problem \eref{eq:risk-guided} is called \emph{front}.

\definecolor{fghblue}{RGB}{0,91,127}
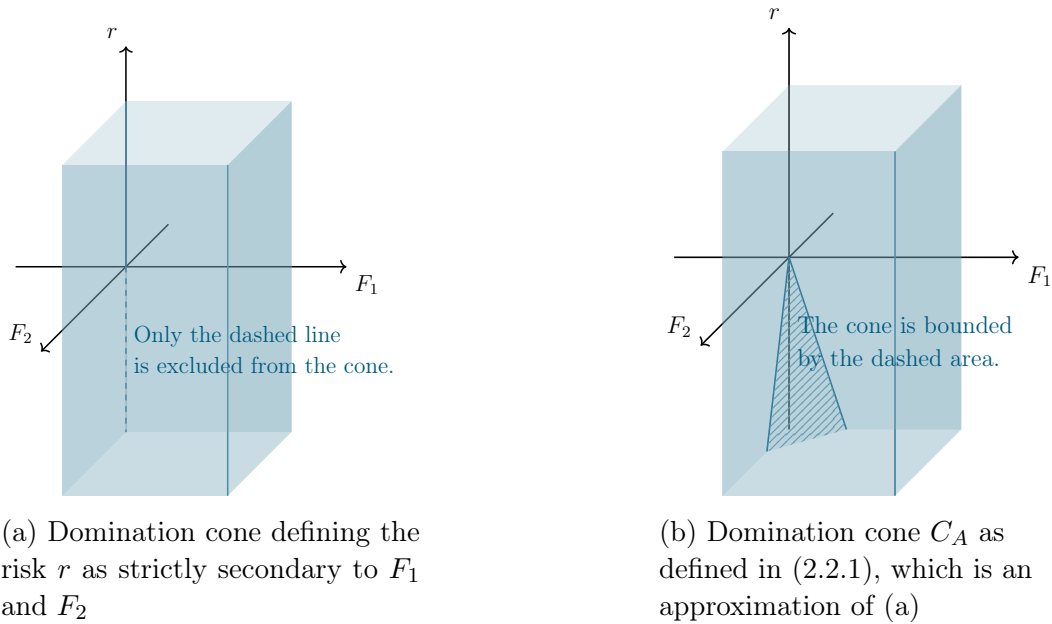
\begin{figure}[h]
 \begin{subfigure}[b]{0.34\textwidth}
  \centering
  \resizebox{\linewidth}{!}{%
  \begin{tikzpicture}
   \begin{scope}[canvas is xy plane at z=0]
    \draw[thick,->] (-2,0) -- (4,0) node[anchor=north west] {$F_1$};
    \draw[thick,->] (0,-3) -- (0,4) node[anchor=south east] {$r$};
\draw[thick, white] (0,0)--(0, -3);
   \end{scope}
   
   \draw[thick,->] (0,0,-2) -- (0,0,4) node[anchor=south east] {$F_2$};
   
   \draw[thick, fghblue!70] (3,3,3)--(3,-3,3);
   \fill[fghblue!70, opacity=0.3] (0,3,3)--(3,3,3) --(3, -3, 3)--(0,-3,3)-- cycle;
   \fill[fghblue!40, opacity=0.3] (0, 3, 3) --(0, 3, 0)--(0, -3, 0)--(0, -3, 3) -- cycle;
   \fill[fghblue!40, opacity=0.3](0, 3, 0)--(3, 3, 0)--(3, -3, 0) -- (0, -3, 0) -- cycle;
   \fill[fghblue!70, opacity=0.3] (3, 3, 0)--(3, -3, 0)--(3, -3, 3) --(3, 3, 3) -- cycle;
   
   \draw[thick, fghblue!70, -] (0,0,0) -- (0,3,0);
   \draw[thick, fghblue!70, dashed] (0,0,0) -- (0,-3,0); 
   \node[align=left, anchor=west, fghblue] at (0,-1.5) {Only the dashed line\\ is excluded from the cone.};
  \end{tikzpicture}}%
  \captionsetup{width=\dimexpr\linewidth+2\fboxsep+2\fboxrule\relax}
        \caption{Domination cone defining the risk $r$ as strictly secondary to $F_1$ and $F_2$}
        \label{fig:lexicographic_cone}
 \end{subfigure}
\hspace{0.2\textwidth}
 \begin{subfigure}[b]{0.34\textwidth}
 \centering
    \captionsetup{justification=raggedright, singlelinecheck=false} 
  \resizebox{\linewidth}{!}{%
  \begin{tikzpicture}
   \begin{scope}[canvas is xy plane at z=0]
    \draw[thick,->] (-2,0) -- (4,0) node[anchor=north west] {$F_1$};
    \draw[thick,->] (0,-3) -- (0,4) node[anchor=south east] {$r$};
   \end{scope}
   
   \draw[thick,->] (0,0,-2) -- (0,0,4) node[anchor=south east] {$F_2$};
   
   \draw[thick, fghblue!80] (3,3,3)--(3,-3,3);
   \fill[fghblue!70, opacity=0.3] (0,3,3)--(3,3,3) --(3, -3, 3)--(0,-3,3)-- cycle;
   \fill[fghblue!40, opacity=0.3] (0, 3, 3) --(0, 3, 0)--(0, 0, 0)--(0, -3, 1)--(0, -3, 3) -- cycle;
   \fill [fghblue!40, opacity=0.3](0, 3, 0)--(3, 3, 0)--(3, -3, 0) -- (1, -3, 0)--(0,0,0) -- cycle;
   \fill [fghblue!40, opacity=0.3](0,0,0)--(1, -3, 0) --(0, -3, 1) --cycle;
   \fill [fghblue!70, opacity=0.3] (3, 3, 0)--(3, -3, 0)--(3, -3, 3) --(3, 3, 3) -- cycle;
   
   \draw[thick, fghblue!80, -] (0,0,0) -- (0,-3,1);
   \draw[thick, fghblue!80, -] (0,0,0) -- (1,-3,0);
    \fill[pattern=north east lines,
        pattern color=fghblue!60]
    (1,-3,0) -- (0,-3,1) -- (0,0,0) -- cycle;
    \node[align=left, anchor=west, fghblue] at (0,-1.5) {The cone is bounded\\ by the dashed area.};
   
  \end{tikzpicture}}%
  \captionsetup{width=\dimexpr\linewidth+2\fboxsep+2\fboxrule\relax}
        \caption{Domination cone $C_A$ as defined in (\ref{eq:CA}), which is an approximation of (a)}
        \label{fig:approximation_cone}
 \end{subfigure}
 \caption{Modeling the risk $r$ as secondary by modifying the domination cone. With the ordering induced by the cone (a), a solution is no longer  optimal if it sacrifices dose objectives for risk reduction. A solution is optimal w.r.t. the ordering induced by the cone $C_A$ (b) if the relative sacrifice in the dose objectives for an improvement in risk $r$ does not exceed a threshold $\epsilon$.}
    \label{fig:cones}
 \end{figure}

\subsubsection{Adjustable parametrization with $\epsilon$}
\label{sec:adj_param_w_eps}

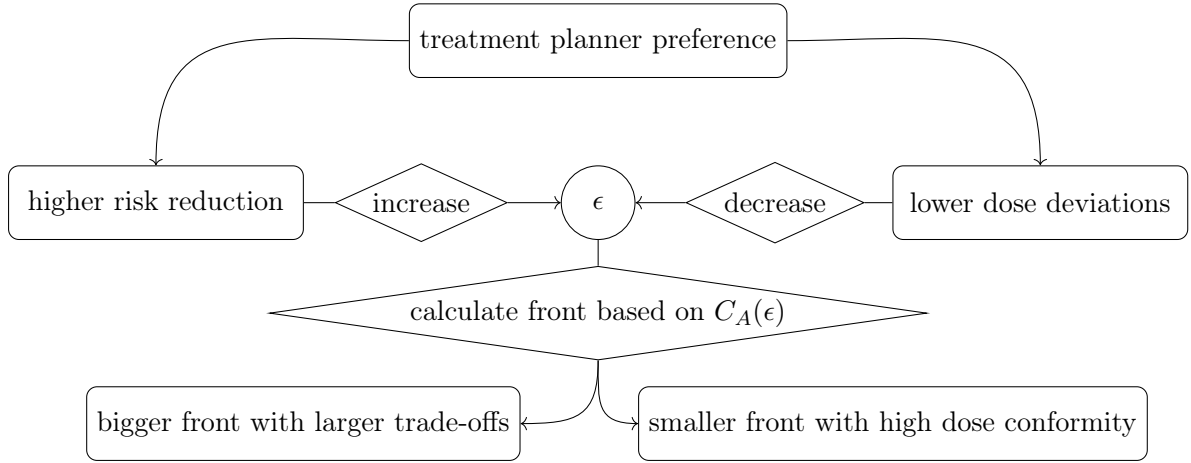
\begin{figure}[h]
    \centering
    \resizebox{\linewidth}{!}{%
\begin{tikzpicture}[font=\small]
  \node[draw,rounded corners,align=center,minimum width=50mm,minimum height=10mm] (tp) at (0,0) {treatment planner preference};
  \node[draw,rounded corners,align=center,minimum width=40mm,minimum height=10mm] (left) at (-6,-2.2) {higher risk reduction};
  \node[draw,rounded corners,align=center,minimum width=40mm,minimum height=10mm] (right) at (6,-2.2) {lower dose deviations};
  \node[draw,circle,align=center,minimum size=10mm] (eps) at (0,-2.2) {$\epsilon$};

  \draw[->] (tp.west) to[out=180,in=90, looseness=1.4] (left.north);
  \draw[->] (tp.east) to[out=0,in=90, looseness=1.4] (right.north);

  \node[draw,shape=diamond,aspect=2.2,align=center,inner sep=2pt,minimum height=8mm] (inc) at (-2.4,-2.2) {increase};
  \node[draw,shape=diamond,aspect=2.2,align=center,inner sep=2pt,minimum height=8mm] (dec) at ( 2.4,-2.2) {decrease};

  \draw (left.east) -- (inc.west);
  \draw[->] (inc.east) -- (eps.west);
  \draw (right.west) -- (dec.east);
  \draw[->] (dec.west) -- (eps.east);

  \node[draw,shape=diamond,aspect=7,align=center,inner sep=2pt,minimum height=10mm] (calc) at (0,-3.7) {calculate front based on $C_A(\epsilon)$};
  \draw (eps.south) -- (calc.north);

  \node[draw,rounded corners,align=center,minimum width=54mm,minimum height=10mm] (big) at (-4.0,-5.2) {bigger front with larger trade-offs};
  \node[draw,rounded corners,align=center,minimum width=54mm,minimum height=10mm] (small) at (4.0,-5.2) {smaller front with high dose conformity};

  \draw[->] (calc.south) to[out=270,in=0, looseness=1.45] (big.east);
  \draw[->] (calc.south) to[out=270,in=180, looseness=1.45] (small.west);
\end{tikzpicture}}
    \caption{Scheme for interactive decision making}
    \label{fig:decission-making}
\end{figure}

The approximated cone $C_A$ employed in the formulation of the problem (\ref{eq:risk-guided}) introduces the parameter $\epsilon$. 
This parameter defines the maximum relative increase in dose objectives that is acceptable for a given reduction in risk. An improvement in the risk by $1 \%$ can be accompanied by a worsening of each dose objective of at most $\frac{\epsilon}{100}$ units. 
The treatment planner can adjust $\epsilon$ to reflect clinical priorities (Fig. \ref{fig:decission-making}). A small $\epsilon$ ensures that the dose objectives are only relaxed when the risk reduction is substantial. A larger $\epsilon$ allows for greater compromises in dose objectives in exchange for moderate risk decreases. For our experiments in Sections \ref{sec:cohort_statistic} and \ref{sec:individual_patient}, we chose $\epsilon$ such that the trade-offs between risk reduction and dosimetric objectives are observable.


Mathematically, shrinking $\epsilon$ enlarges the domination cone $C_A$. As a result, more solutions become dominated, and the set of non-dominated (optimal) solutions contracts. In the limit $\epsilon \rightarrow 0$, the only permissible risk improvements are those that do not worsen any dose objective. Conversely, as $\epsilon \rightarrow \infty$  the prioritization between dose objectives and risk disappears, and the resulting plans coincide with those produced by the conventional \ac{MCO} formulation, where the risk is treated equally to the dose objectives. In \ref{sec:contrast_to_conventional} we investigate the difference from conventional MCO in more detail. In \ref{sec:adjustable_parametrization}, we demonstrate how different choices of the parameter $\epsilon$ affect the calculated fronts. 

\subsubsection{Approximation of the front for the risk-including problem}

The Sandwiching algorithm described in Section \ref{sec:POplans} can be used to calculate the front of any \ac{MCO} problem for which the order is induced by a polyhedral domination cone. As $C_A$ is indeed a polyhedral cone, we can apply Sandwiching to efficiently calculate a good representation of the front for problem (\ref{eq:risk-guided}), much in the same way as for the risk-agnostic problem (\ref{eq:risk-agnostic-MCO}). 


%% file: Results.tex


In the following, we evaluate our proposed method both on a cohort level and a patient specific level. For our cohort-wide analysis, we examined 19 patients from the NSCLC data set described in Section \ref{sec:data}. For all 19 patients, we calculated the fronts for both the risk-agnostic model (Section \ref{sec:risk_agnostic_model}) and our proposed model with RP risk as secondary objective (Section \ref{sec:integrating_the_risk_model}). The  dose objectives and constraints used are detailed in Section \ref{sec:dosimetric_model}, while the risk model is described in \ref{sec:risk-model} and \ref{sec:appendix_risk_model}. Based on the results of these calculations, we show exemplary results for our method (Section \ref{sec:inter_patient_var}) and provide a cross-cohort statistic (Section \ref{sec:cohort_statistic}.) The statistic is based on a paired comparison between a risk-agnostic and risk-guided plan, which were systematically chosen as detailed in \ref{sec:measuring}.
For an individual patient, we demonstrate the effect of choosing different values for the parameter $\epsilon$ (Section \ref{sec:adjustable_parametrization}), highlight the difference from conventional MCO (Section \ref{sec:contrast_to_conventional}), and display DVH curves and dose color washes (Section \ref{sec:individual_patient}). We also exemplarily apply our method to three dose objectives instead of two (Section \ref{sec:three_objectives}). 

For the calculation of the plan sets, an in house Sandwiching algorithm implementation by Fraunhofer ITWM Optimization department was used. The scalarized optimization problems were implemented in Python 3.12 and solved with the commercially available knitro solver \cite{ByrNocWal06} on a Lenovo T490s laptop with a 13th Gen Intel Core i7-1370P 1.90 GHz processor. For dose calculation and image creation, we employed \emph{matRad} \cite{CisMaiZie15}, a MATLAB \cite{MATLAB} and GNU Octave \cite{eaton:2002} based open source software for radiation treatment developed by the German Cancer Research Center.


\subsection{Experiment design}

\subsubsection{Patient data}
\label{sec:data}
Our methodology is built and tested on a retrospective data set of stage IIB-IIIA \ac{NSCLC} patients treated with proton (passive scattering proton therapy, PSPT) or photon (intensity-modulated radiotherapy, IMRT) radiation. Patients were treated with radiation doses of 60-74 Gy (accounting for relative biologic effectiveness [RBE] of 1.1) in 1.8-2 Gy/fractions and received concurrent chemotherapy. Our main toxicity endpoint on which the risk model was built was the incidence of radiation pneumonitis of grade 2 or higher. 

\subsubsection{Dosimetric model}
\label{sec:dosimetric_model}
When optimizing treatment plans, we employ two dose-based objectives for the heart and spinal cord, as well as a total of nine dose-based constraints for \ac{PTV}, heart, spinal cord, total lung, and esophagus. The dose limits imposed by the constraints were chosen based on the clinical prescriptions provided by MGH.

Table \ref{tab:optimization_model} details how we used these functions as objectives and constraints in the optimization model. For patients in whom the common model proved insufficient, manual adjustments were made.
\begin{table}[ht!]
	\small
    \caption{The optimization problem. Reference values are given in Gy.}
	\begin{center}
		\begin{tabular}{l | l | l } 
			\hline
			\textbf{Volume} & \textbf{Evaluation function} & \textbf{Parameters} \\
			\hline
			\multicolumn{3}{c} {\rule{0pt}{4ex}  Objectives}  \\ 
			\hline
			Heart & EUD  & $p=1$\\
			Spinal cord & EUD & $p=2$ \\
			\hline
			\multicolumn{3}{c} {\rule{0pt}{4ex} Constraints}  \\ 
			\hline
			Heart & EUD & $p=1$, \qquad \qquad ~$\leq 30$ \\
            Heart & OD & $p=2$, $\refDose = 60$, $\leq 0$ \\
            Total lung & V & \quad ~~~~~ $\refDose = 20$, $\leq 12\%$ \\
            Total lung & OD & $p=2$, $\refDose = 30$, $\leq 0$\\
            PTV & UD & $p=2$, $\refDose = 60$, $\leq 0.15$ \\
            PTV & OD & $p=2$, $\refDose = 66$, $\leq 0$ \\
			Esophagus & V & \quad ~~~~~ $\refDose =55$, $\leq 20\%$ \\
            Esophagus & OD & \quad ~~~~~ $\refDose = 60$, $\leq 0$ \\
            Spinal cord & OD & $p=2$, $\refDose = 50$, $\leq 0$\\
            \hline
		\end{tabular}
	\end{center}
    \label{tab:optimization_model}
\end{table}

\subsubsection{Risk model for radiation pneumonitis}
\label{sec:risk-model}
To predict the risk of our primary endpoint (\ac{RP} grade $\geq 2$, binary endpoint), we trained a logistic regression on a cohort of 69 \ac{NSCLC} patients, containing dosimetric, clinicopathological (age, sex, smoking status, breathing function), and biological data on pre- and mid-treatment (week ~4-6) [18]F-flurodeoxyglucose (FDG)-PET scans as an indicator of normal lung inflammation. More details about the cohort are given in our previous publication \cite{Ajdari_2022}. A bootstrapped stepwise logistic regression allowing for interaction terms was trained on the cohort. Starting from a null (intercept only) model, features were iteratively added if their inclusion improved model's fit, as assessed by the Bayesian Information Criterion (BIC). The highest predictive performance, in terms of area under the receiver operating characteristic curve (AUROC) was achieved for a parsimonious model with only four risk modifiers: dosimetric factors (V$\left[5Gy\right]$ for the right lung and V$\left[20Gy\right]$ for the total lung $(TL)$), and two non-dosimetric modifiers: current smoking status (binary, currently smoking or not), and baseline breathing function (forced exhalation volume capacity, in percentage). 
More information on the model, including the bootstrapped-ROC curve and the calibration plot along with the predicted risk of RP for all patients in the cohort, can be found in Appendix \ref{sec:appendix_risk_model}.

\subsubsection{Measuring the benefit of risk-guided planning}
\label{sec:measuring}
To assess the potential benefit of our approach, for each patient in the cohort, we performed a paired comparison between two plans: A first treatment plan optimized solely on dosimetric objectives (risk-agnostic plan) and a second plan selected from the risk-guided plan set produced by our method (risk-guided plan). 

For the risk-agnostic plan, we chose the plan for which both objectives were weighted similarly, i.e., the result of the scalarization problem (\ref{eq:weighted_sum_scal}) with $w_{1} \approx w_{2}$. This reflects a planner for whom both objectives are roughly equally important.
 

To select the risk-guided plan for each patient, we first excluded any plan whose dose objectives deviated by more than ±$30\%$ of the overall objective range from those of the risk-agnostic plan. From the remaining plans, we then — whenever possible — retained only those satisfying the prescribed trade-off $\epsilon = 10$ between dose degradation and risk reduction (i.e., true re-optimizations of the risk-agnostic plan). If no such plan existed, a possibility due to the finite set of calculated plans, we kept all plans within the $30\%$ window. Finally, we chose the candidate with the lowest predicted risk for the paired comparison. Figure \ref{fig:risk-guided-selection} visualizes this selection process of plans to compare.

\newlength{\narrowboxwidth}
\setlength{\narrowboxwidth}{3.6cm} 
\newlength{\decisionwidth}
\setlength{\decisionwidth}{3.2cm}  
\newlength{\discardgap}
\setlength{\discardgap}{14mm}      
\tikzset{
  block/.style={rectangle, draw, rounded corners, align=left, inner sep=6pt, minimum width=30mm},
  narrowblock/.style={rectangle, draw, rounded corners, align=left, inner sep=1pt, text width=\narrowboxwidth, minimum height=8mm},
  decision/.style={diamond, draw, align=center, inner sep=1pt, text width=\decisionwidth, aspect=2},
  line/.style={-Latex, thick}
}
\begin{figure}
    \centering
    \resizebox{\linewidth}{!}{\begin{tikzpicture}[node distance=7mm and 9mm]
  \node[block] (start) {Start: \\- risk-agnostic plan RA\\- risk-guided front};

  \node[rectangle, draw, rounded corners, align=left, inner sep=6pt, text width=3.2cm, right=of start] (discard_dose) {Discard risk-guided plans with dose objective values too different from RA};

  \node[decision, right=of discard_dose] (tradeoff) {Are there risk-guided plans with lower relative trade-off than $\epsilon$ from RA?};

  \node[block, right=of tradeoff] (select) {Select plan with lowest risk};

  \node[block, below=\discardgap of select] (discard_others) {Discard other plans};

  \draw[line] (start) -- (discard_dose);
  \draw[line] (discard_dose) -- (tradeoff);
  \draw[line] (tradeoff) -- node[above]{No} (select);
  \draw[line] (tradeoff) |- node[pos=0.25,left]{Yes} (discard_others);
  \draw[line] (discard_others) -- (select);
\end{tikzpicture}}
    \caption{Scheme of select a risk-guided plan for comparison.}
    \label{fig:risk-guided-selection}
\end{figure}
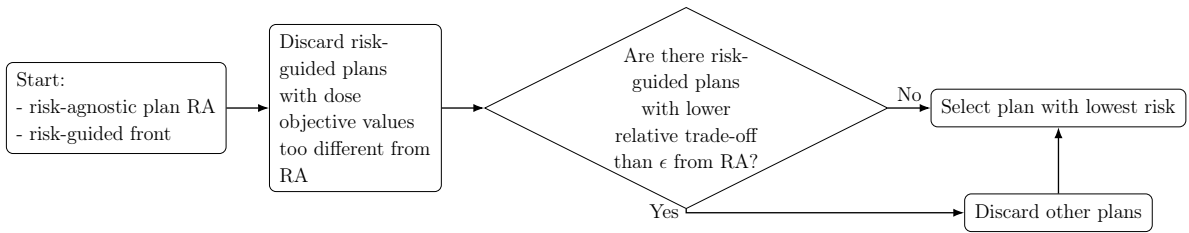

\subsection{Experiments}

\subsubsection{Exemplary outcomes for risk-guided \ac{MCO}}
\label{sec:inter_patient_var}

\begin{figure}[h] 
\centering 
\begin{subfigure}[b]{0.45\textwidth} 
    \centering \includegraphics[width=\linewidth]{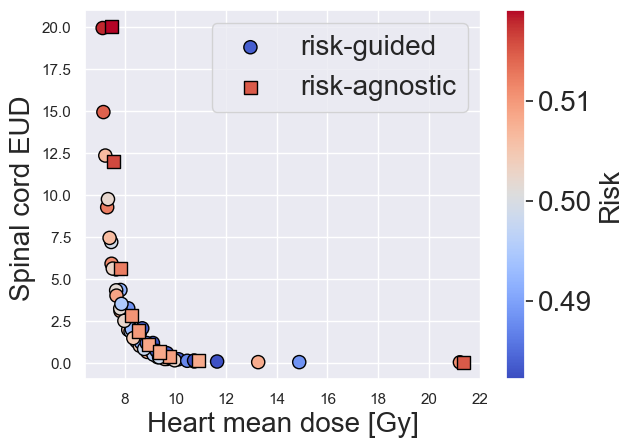}
    \caption{Patient for whom only small risk-improvement is possible}
\end{subfigure} 
\hfill
\begin{subfigure}[b]{0.45\textwidth} 
    \centering 
    \includegraphics[width=\linewidth]{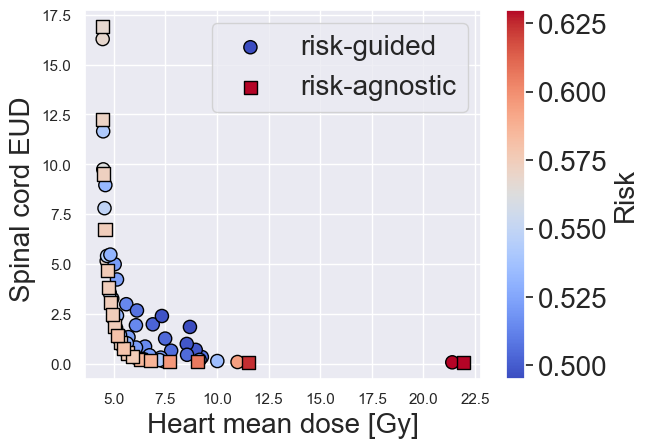}
    \caption{Patient for whom risk-improvement is possible at a cost in the dose objectives.}
    \end{subfigure} 
\caption{Different possibilities to improve risk prediction and to deviate from dose objectives for different patients.} \label{fig:types_of_risk_improvement}
\end{figure}

Figure \ref{fig:types_of_risk_improvement} shows the risk-agnostic and risk-guided fronts for three representative patients, highlighting both the achievable risk reduction with risk-guided plan optimization and the degree of deviation from the risk-agnostic front. For the calculation of the risk-guided fronts $\epsilon = 10$ was chosen in each case.

By design, our method omits any risk-guided plan that provides only a marginal risk reduction, yet lies far from the risk-agnostic front. Depending on the patient, it is possible that nearly no acceptable dose trade-offs lead to a sufficient risk-improvement (Fig.\ref{fig:types_of_risk_improvement}(a)), or that the risk can be reduced by deviation slightly from dose optimal plans (Fig.\ref{fig:types_of_risk_improvement}(b)).

\subsubsection{Adjustable parametrization of our method}
\label{sec:adjustable_parametrization}
As discussed in Section \ref{sec:adj_param_w_eps}, the parameter $\epsilon$ in the algorithm for calculating the risk-guided front defines the maximum relative increase in dose objectives that is acceptable for a given reduction in risk. 

\begin{figure}[h]
    \centering
     \begin{subfigure}[b]{0.32\textwidth}
        \centering
        \includegraphics[width=\textwidth]{figures/175_eps10_combined.png}
        \caption{$\epsilon=10$}
        \label{fig:large_epsilon}
    \end{subfigure}
    \hfill
    \begin{subfigure}[b]{0.32\textwidth}
        \centering
        \includegraphics[width=\textwidth]{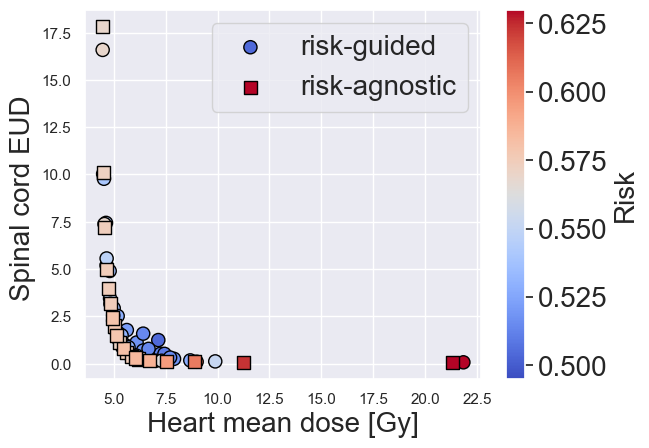}
        \caption{$\epsilon=5$}
        \label{fig:medium_epsilon}
    \end{subfigure}
    \hfill
        \begin{subfigure}[b]{0.32\textwidth}
        \centering
        \includegraphics[width=\textwidth]{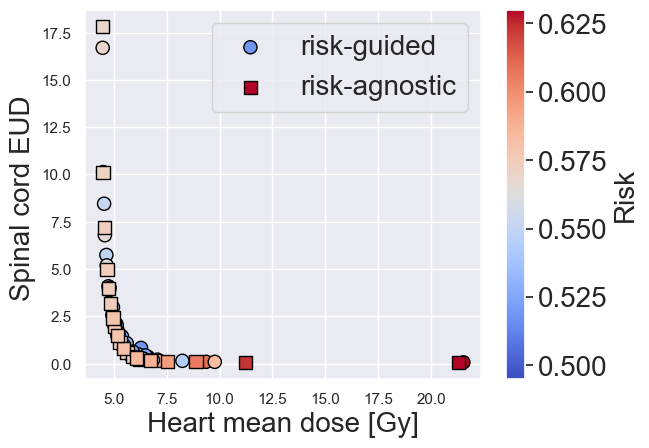}
        \caption{$\epsilon = 2.5$}
        \label{fig:small_epsilon}
    \end{subfigure}
    \caption{Risk-agnostic and risk-guided fronts calculated with three different values for $\epsilon$ for a representative patient (Patient 19 from Table \ref{tab:statistics}).}
    \label{fig:different_epsilon}
\end{figure}

\begin{figure}[h]
    \centering
        \centering
        \includegraphics[width=0.6\textwidth]{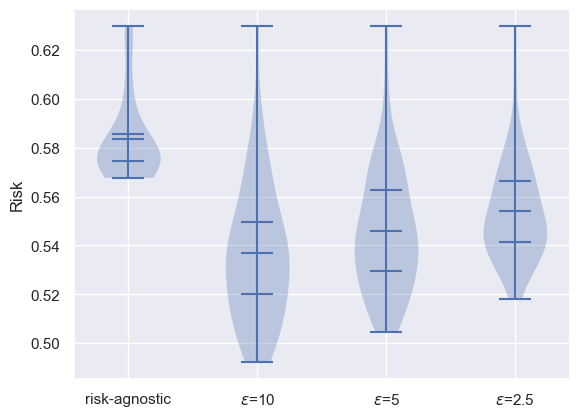}
    \caption{Frequency distribution for a representative patient (Patient 19 from Table \ref{tab:statistics}): The plots show the risk-values of the representative sets for the fronts calculated with the risk-agnostic approach and the risk-guided approach with $\epsilon=10$, $\epsilon = 5$, and $\epsilon = 2.5$. The horizontal lines represent the risk quantiles (min, 25\%, mean, 75\%, and max).}
    \label{fig:distribution_epsilon}
\end{figure}

Figure \ref{fig:different_epsilon} shows the resulting fronts for one of the patients for three different values of $\epsilon$: $2.5$, $5$ and $10$. Figure \ref{fig:distribution_epsilon} shows the frequency distribution of the risk-values of the representative sets of the Pareto fronts calculated for different values of $\epsilon$. The example demonstrates how adjustment of $\epsilon$ allows the planner to decide whether to explore a wide range of risk to dose trade-offs, or whether to focus more strictly on minimizing dose objectives.

\subsubsection{In-depth analysis of an individual patient} 
\label{sec:individual_patient}

\begin{figure}
\centering
\begin{minipage}[t]{0.45\textwidth}
    \vspace{0pt}
    \centering
    \subcaptionbox{Pareto fronts of the risk-agnostic and risk-guided approach with selected plans for detailed analysis.\label{fig:selected_plans}}{%
        \includegraphics[width=\textwidth]{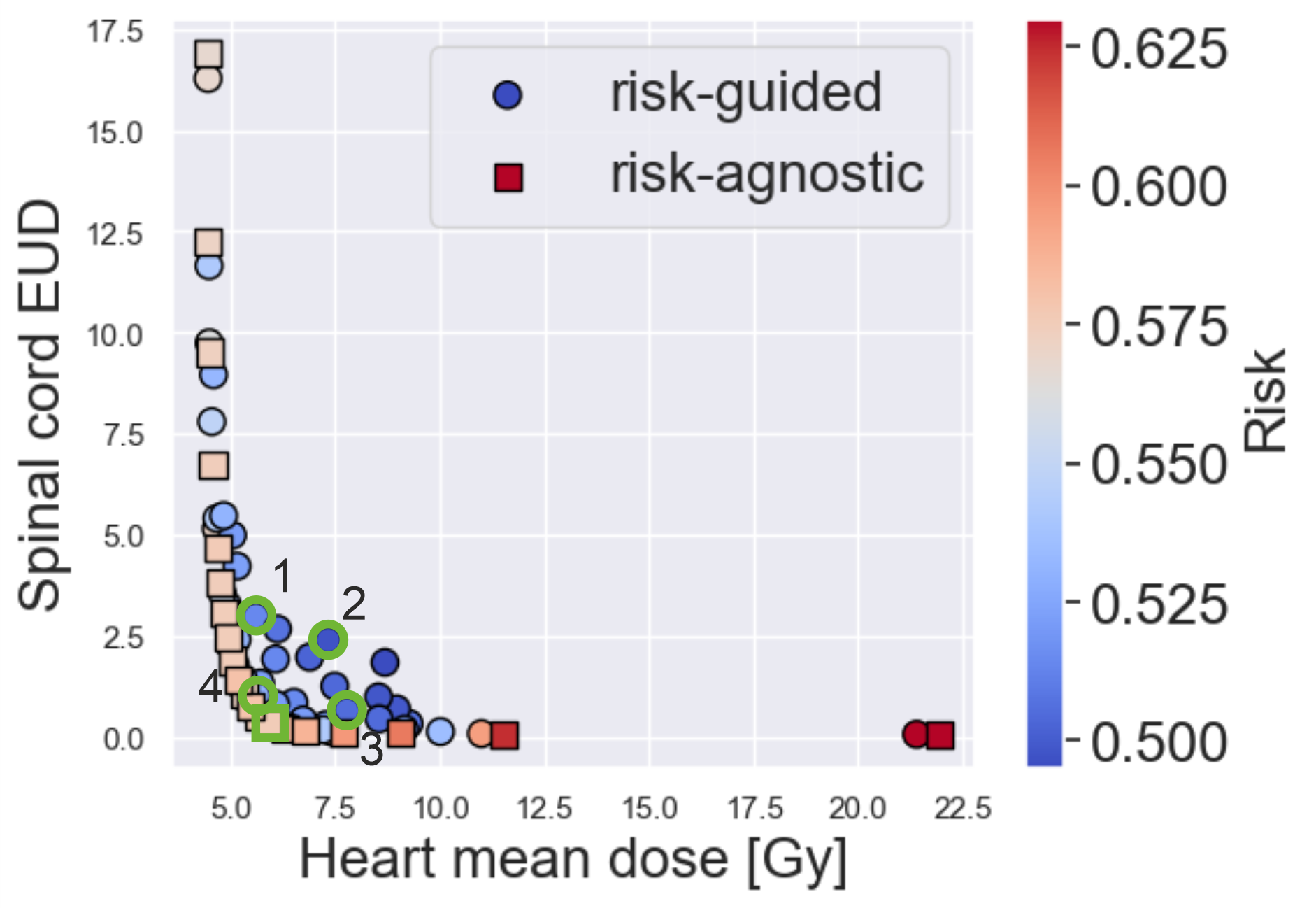}%
    }
\end{minipage}
\hfill
\begin{minipage}[t]{0.45\textwidth}
    \vspace{0pt}
    \centering
    \includegraphics[width=0.5\textwidth]{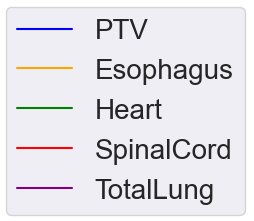}
\end{minipage}
    \\
    \begin{subfigure}[b]{0.45\textwidth}
    
        \centering
        \includegraphics[width=\textwidth]{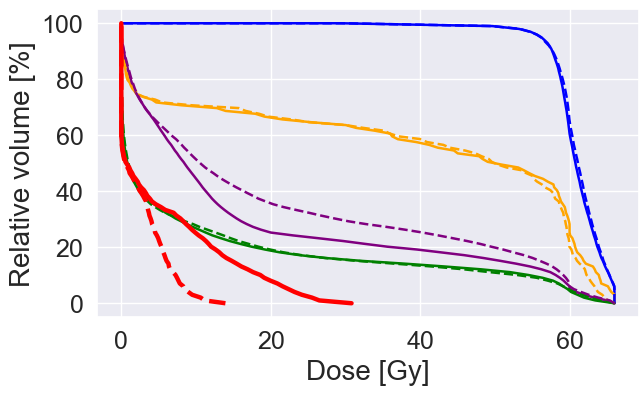}
        \caption{DVHs of the risk-agnostic plan and a risk-guided plan with trade-offs in the spinal cord (Plan 1).}
        \label{fig:DVH_difference_spincalcord}
    \end{subfigure}
    \hfill
     \begin{subfigure}[b]{0.45\textwidth}
        \centering
        \includegraphics[width=\textwidth]{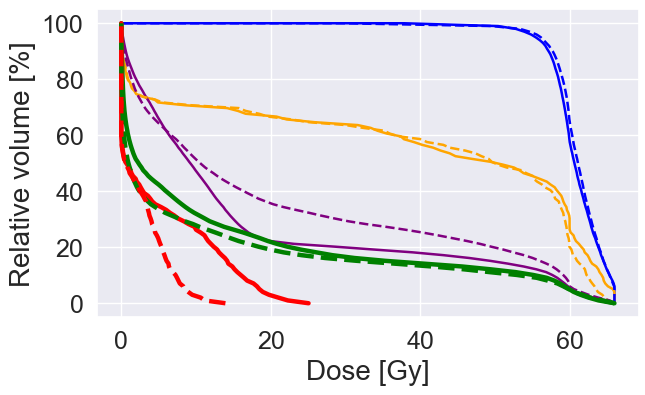}
        \caption{DVHs of the risk-agnostic plan and a risk-guided plan with trade-offs in the spinal cord and heart (Plan 2).}
        \label{fig:DVH_difference_both}
    \end{subfigure}
    \vspace{1cm} 
     \begin{subfigure}[b]{0.45\textwidth}
        \centering
        \includegraphics[width=\textwidth]{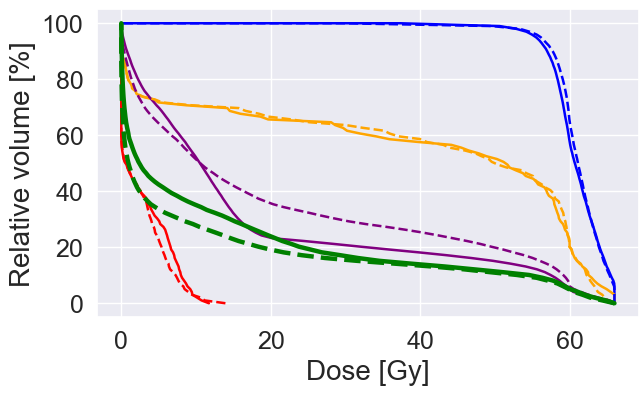}
        \caption{DVHs of the risk-agnostic plan and a risk-guided plan with trade-offs in the heart (Plan 3).}
        \label{fig:DVH_difference_heart}
    \end{subfigure}
    \hfill
     \begin{subfigure}[b]{0.45\textwidth}
        \centering
        \includegraphics[width=\textwidth]{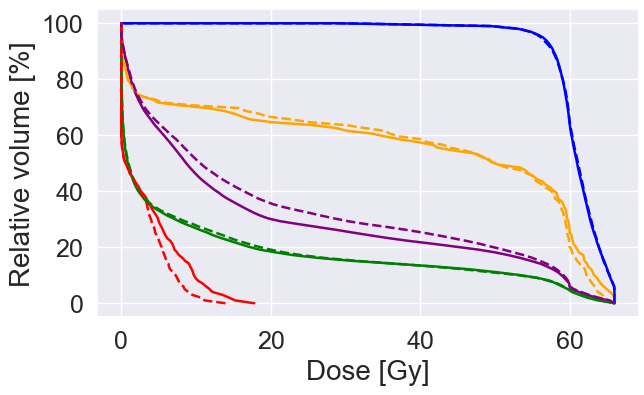}
        \caption{DVHs of the risk-agnostic plan and a risk-guided plan with  only minimal trade-offs (Plan 4).}
        \label{fig:DVH_difference_minimal}
    \end{subfigure}
    \caption{Comparison of a risk-agnostic plan with risk-guided plans focusing on different trade-offs. Solid lines belong to the risk-guided plan, dashed-lines to the risk-agnostic plans.}
    \label{fig:difference_risk_guided_agnostic}
\end{figure}

In this section, we focus on a single patient to assess how the risk-guided approach alters the treatment plan. We computed both the risk-agnostic and risk-guided plan sets, using $\epsilon = 10$ for the latter. From the depiction of the plan sets in Figure \ref{fig:difference_risk_guided_agnostic}(a) we see that, for this patient, a substantial reduction in risk is achievable.

We then picked risk agnostic plans and performed a paired comparison between those plans and a good candidate plan from the risk guided plan set. The selected plans are highlighted in \ref{fig:difference_risk_guided_agnostic}(a). 

Figures \ref{fig:difference_risk_guided_agnostic}(b)-(e) show the \ac{DVH} curves for both the selected risk-agnostic plan and one selected risk-guided plan. Figure \ref{fig:difference_risk_guided_agnostic_dosewash} shows the respective dose distributions for one slice per case and the difference to the risk-agnostic plan. The risk-guided plans were chosen to show the variety of trade-offs that are possible to achieve a risk reduction. Figure \ref{fig:difference_risk_guided_agnostic}(b) and Figure \ref{fig:difference_risk_guided_agnostic_dosewash}(b) and (c) display a plan in which a trade-off  was accepted in the spinal cord.  In our model we ensured with constraints that the dose to the spinal cord will not exceed 45 Gy, and even though there is a dose increase it stays well away from this limit. Figure \ref{fig:difference_risk_guided_agnostic}(c) and Figures \ref{fig:difference_risk_guided_agnostic_dosewash}(d) and (e) belong to a plan with trade-offs in both objectives, Figure \ref{fig:difference_risk_guided_agnostic}(d), \ref{fig:difference_risk_guided_agnostic_dosewash}(f) and (g) display the case where the trade-off is mainly in the heart objective. Finally, Figure \ref{fig:difference_risk_guided_agnostic}(e) and Figures \ref{fig:difference_risk_guided_agnostic_dosewash}(h) and (i) show the results for a plan chosen close to the comparable risk-agnostic plan. All plans share that the risk reduction is achieved by a lower dose to the lung, as this is one of the main contributing factors according to our risk model. Small changes in the dose are observable in other structures. This reflects the fact that we are willing to find a completely new plan as long as the dose objectives remain similar.

\newlength{\SubFigMaxH}
\setlength{\SubFigMaxH}{2.5cm}
\begin{figure}
    \centering
     \begin{subfigure}[b]{0.45\textwidth}
        \centering
        \includegraphics[width=\textwidth,height=\SubFigMaxH,keepaspectratio]{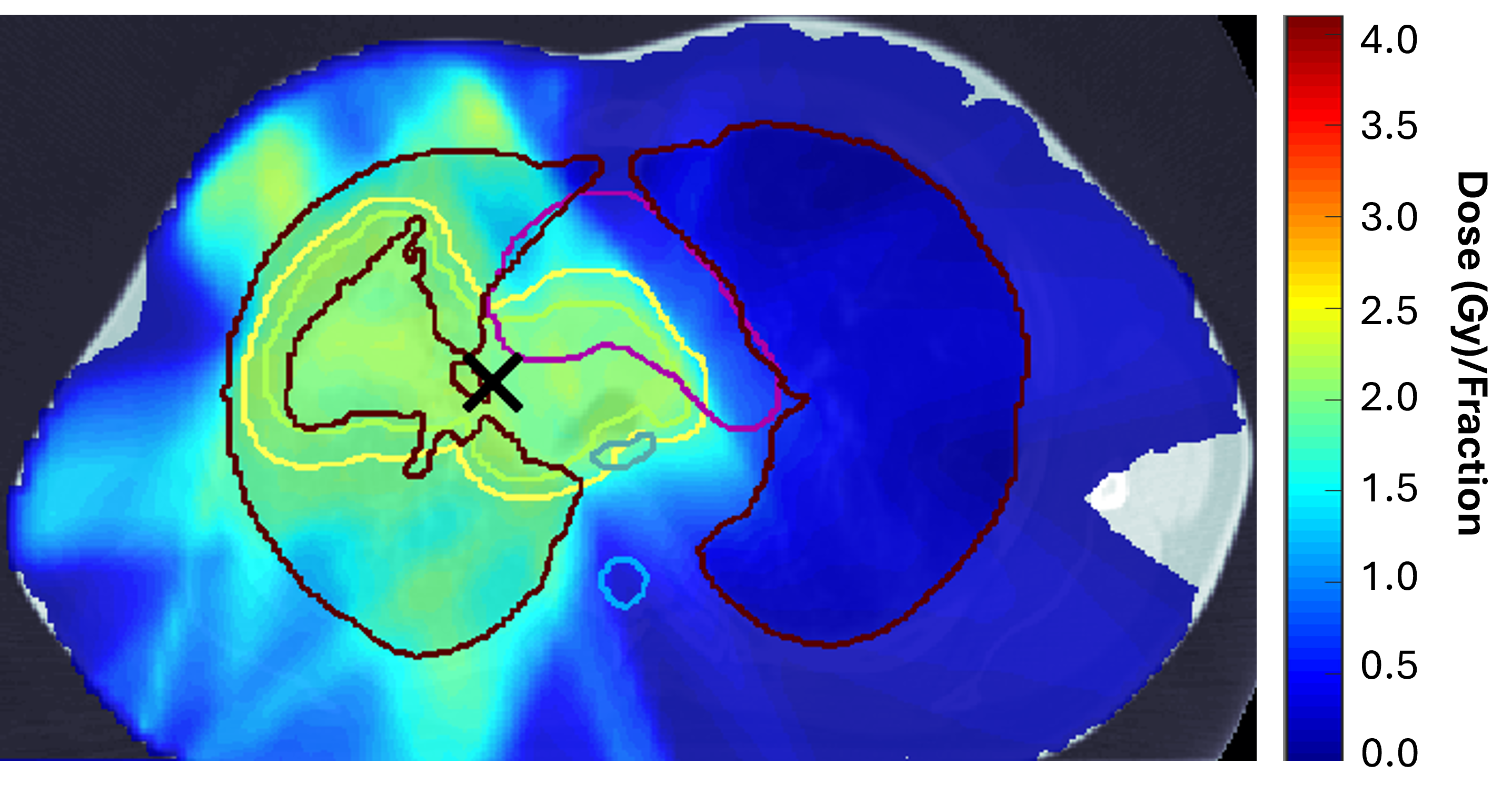}
        \caption{Risk-agnostic plan}
    \end{subfigure}
    \\
    \begin{subfigure}[b]{0.45\textwidth}
        \centering
        \includegraphics[width=\textwidth,height=\SubFigMaxH,keepaspectratio]{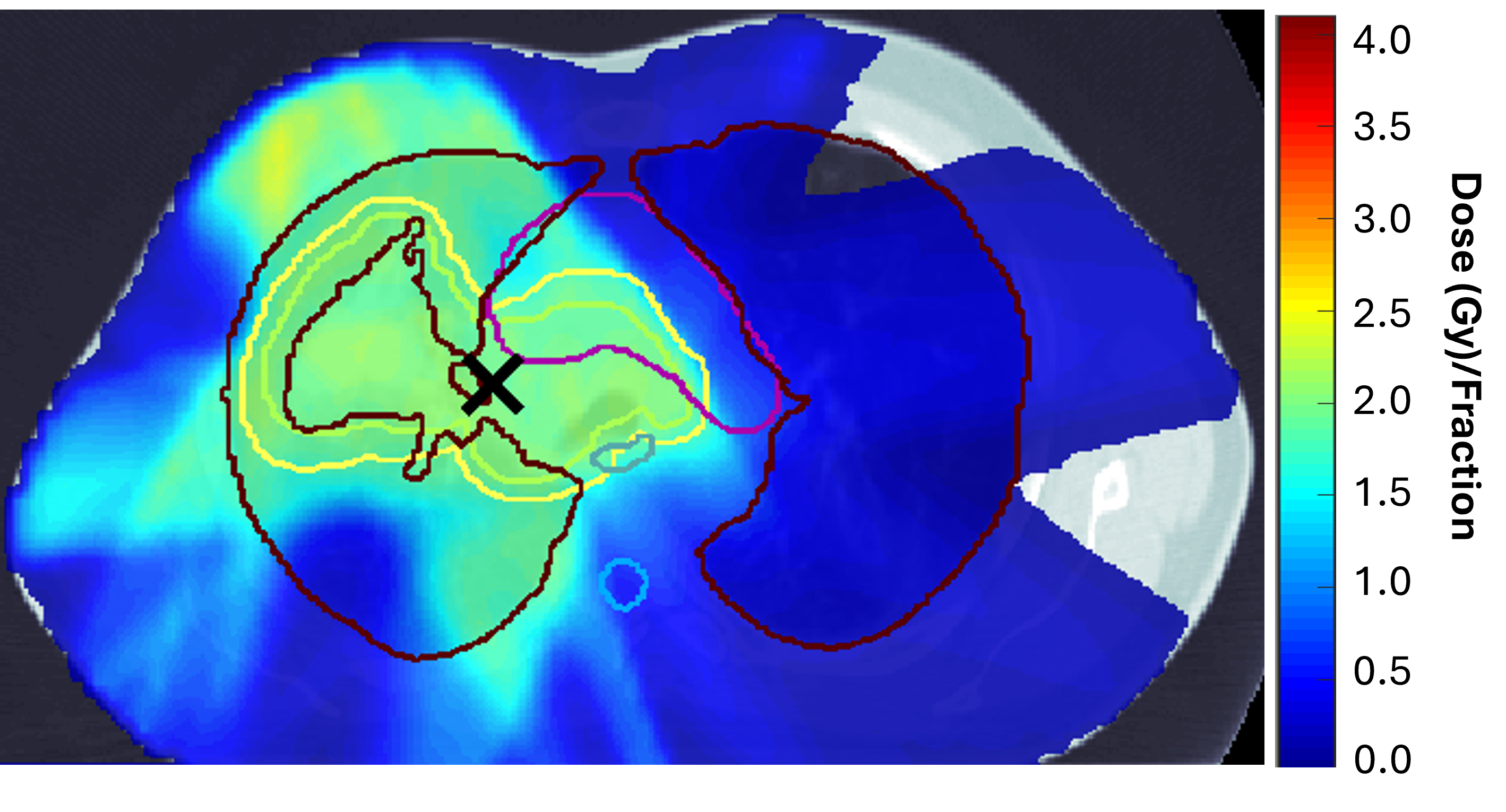}
        \caption{Risk-guided plan with trade-offs in the spinal cord (Plan 1).}
    \end{subfigure}
    \hfill
    \begin{subfigure}[b]{0.45\textwidth}
        \centering
        \includegraphics[width=\textwidth,height=\SubFigMaxH,keepaspectratio]{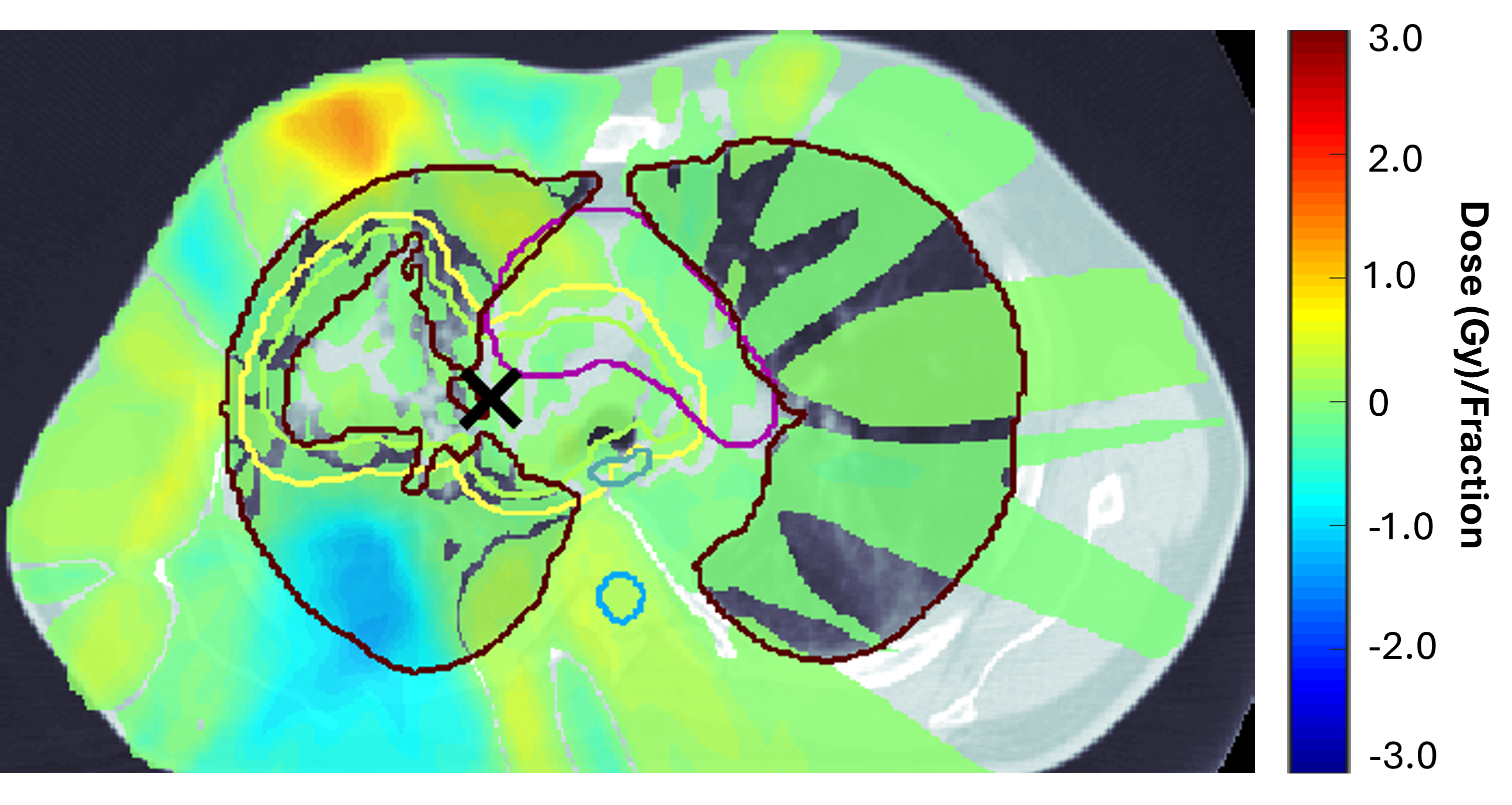}
        \caption{Difference between Plan 1 and the risk-agnostic plan}
    \end{subfigure}
    \\ 
     \begin{subfigure}[b]{0.45\textwidth}
        \centering
        \includegraphics[width=\textwidth,height=\SubFigMaxH,keepaspectratio]{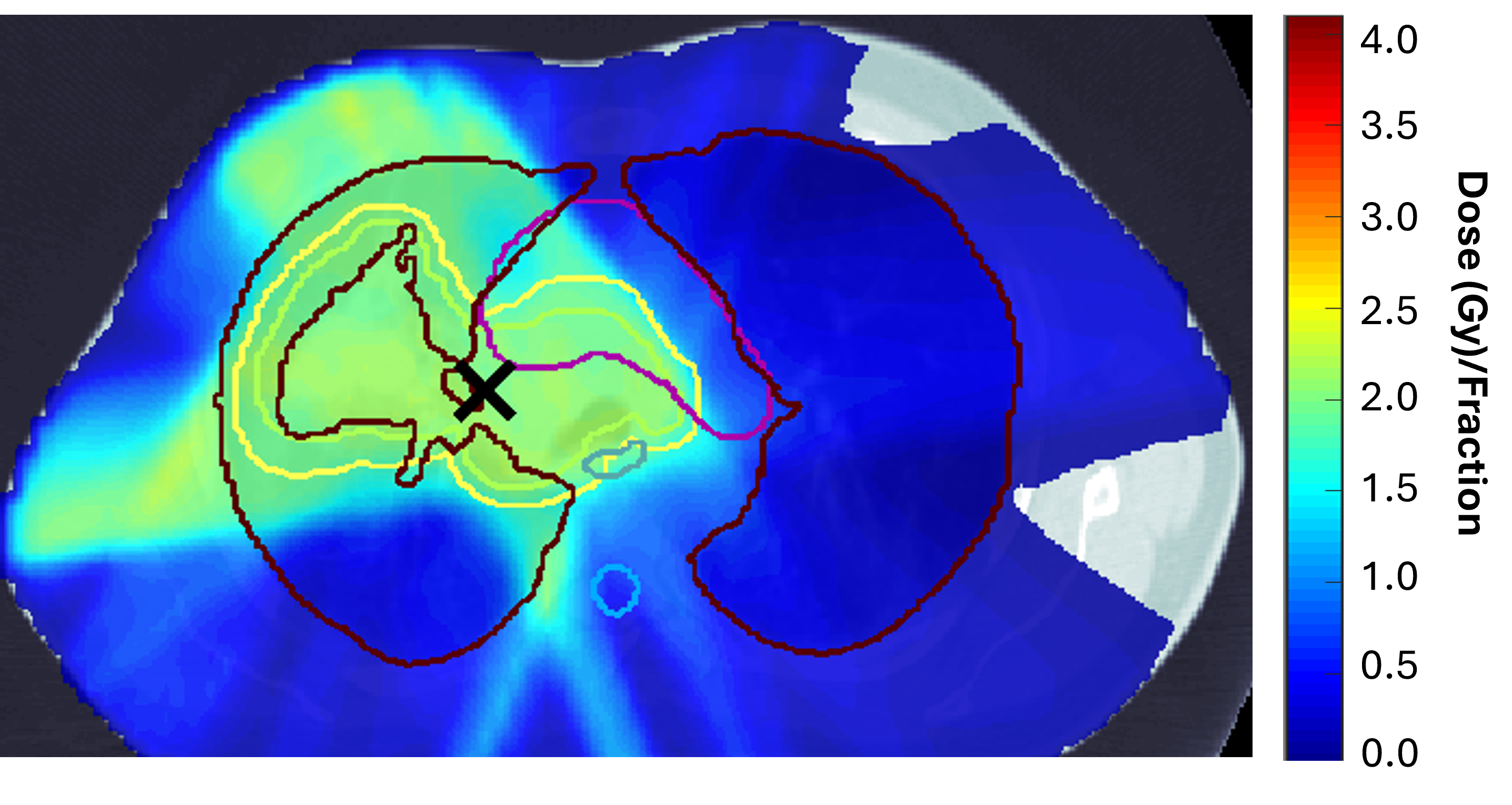}
        \caption{Risk-guided plan with trade-offs in the spinal cord and heart (Plan 2).}
    \end{subfigure}
    \hfill
    \begin{subfigure}[b]{0.45\textwidth}
        \centering
        \includegraphics[width=\textwidth,height=\SubFigMaxH,keepaspectratio]{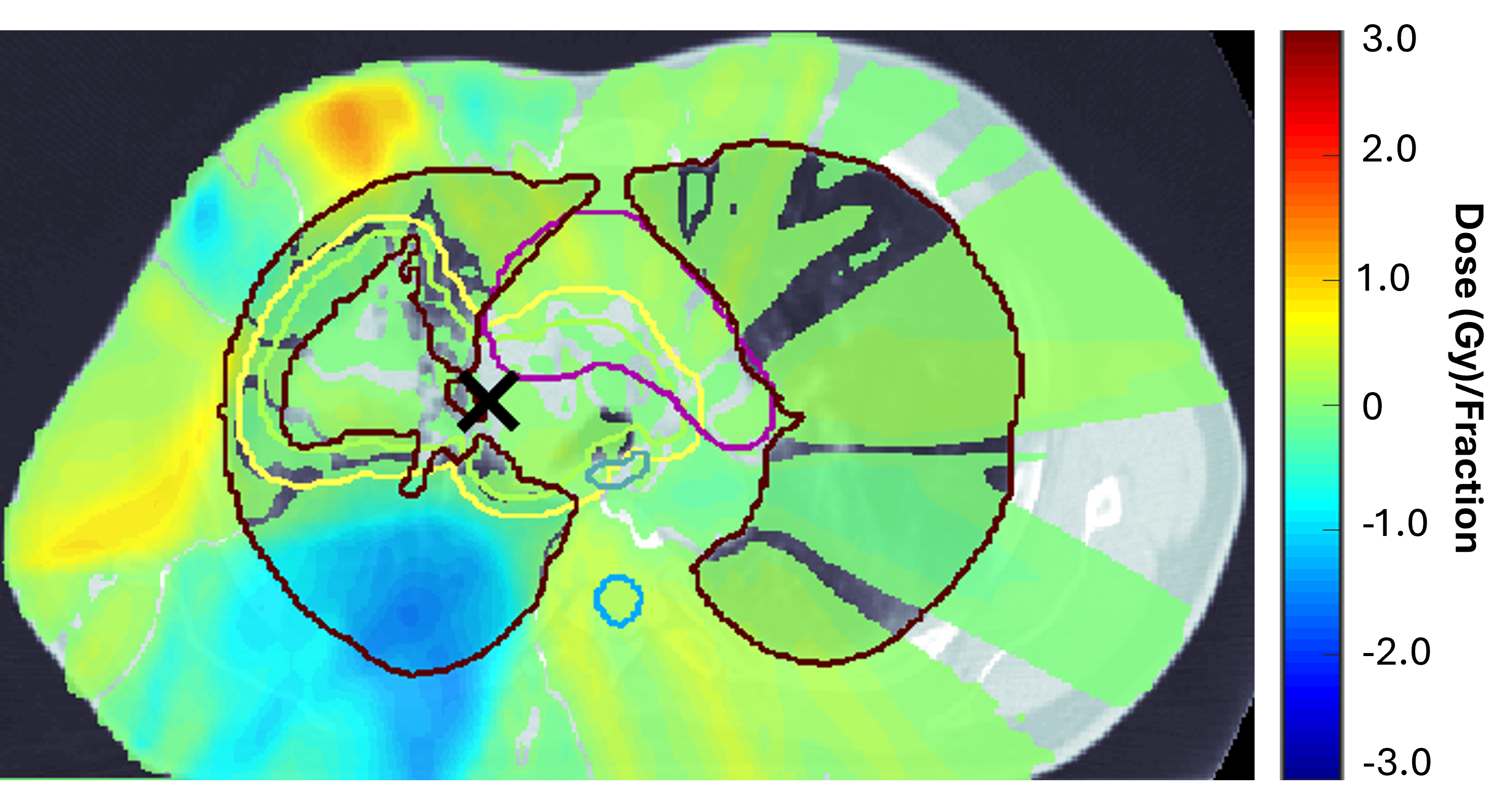}
        \caption{Difference between Plan 2 and the risk-agnostic plan}
    \end{subfigure}
    \\
     \begin{subfigure}[b]{0.45\textwidth}
        \centering
        \includegraphics[width=\textwidth,height=\SubFigMaxH,keepaspectratio]{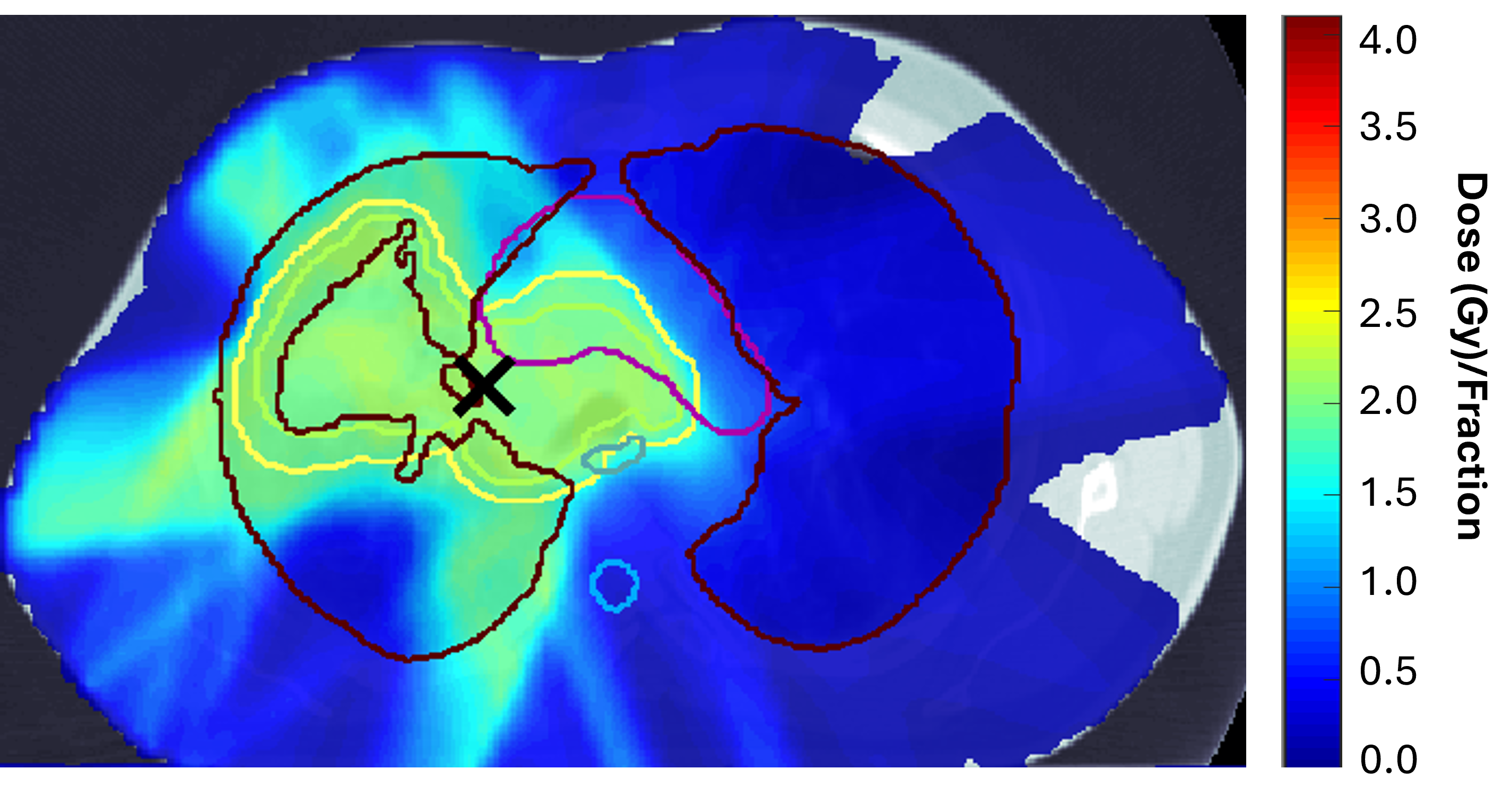}
        \caption{Risk-guided plan with trade-offs in the heart (Plan 3).}
    \end{subfigure}
    \hfill
    \begin{subfigure}[b]{0.45\textwidth}
        \centering
        \includegraphics[width=\textwidth,height=\SubFigMaxH,keepaspectratio]{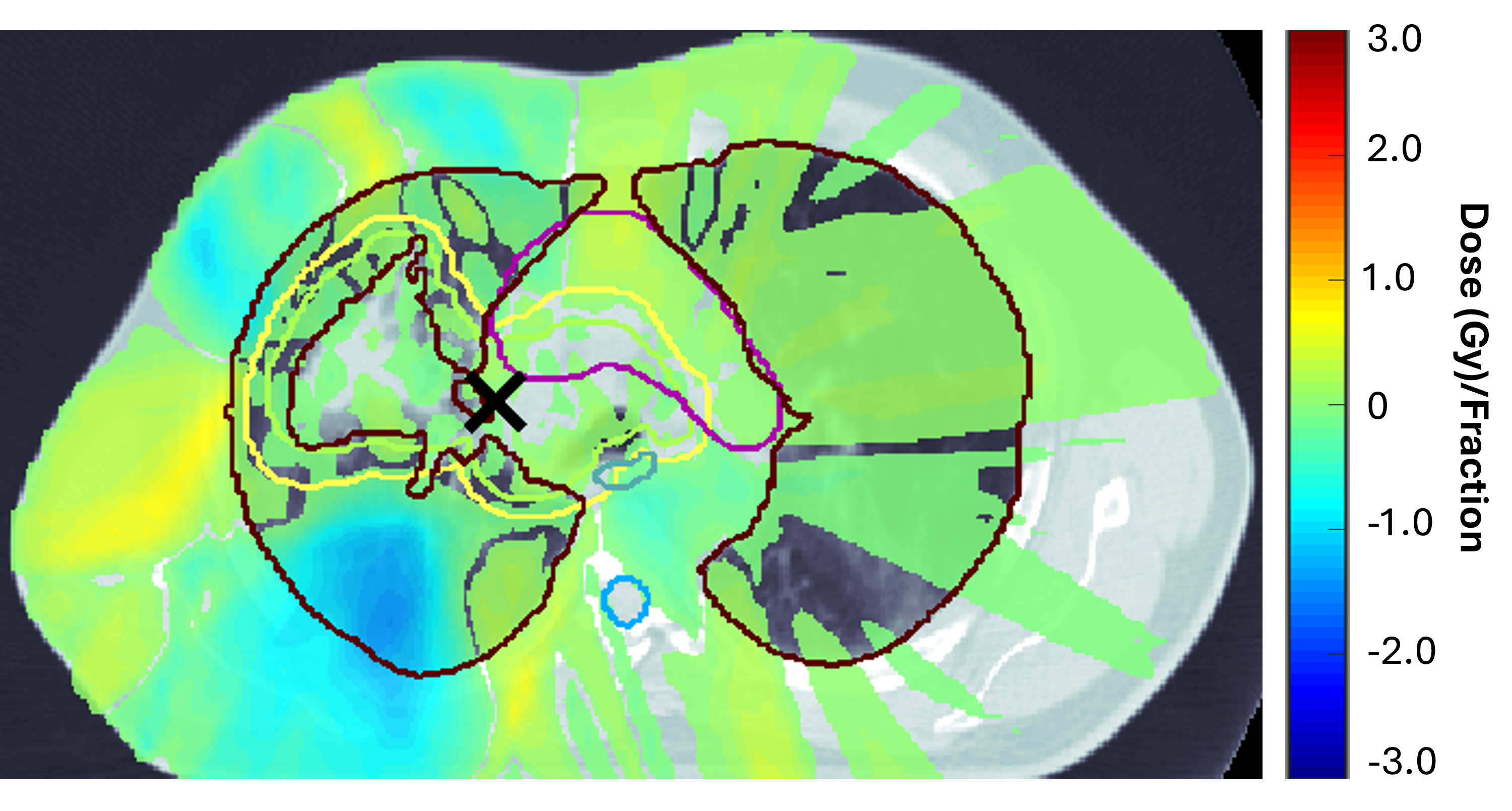}
        \caption{Difference between Plan 3 and the risk-agnostic plan}
    \end{subfigure}
    \\
     \begin{subfigure}[b]{0.45\textwidth}
        \centering
        \includegraphics[width=\textwidth,height=\SubFigMaxH,keepaspectratio]{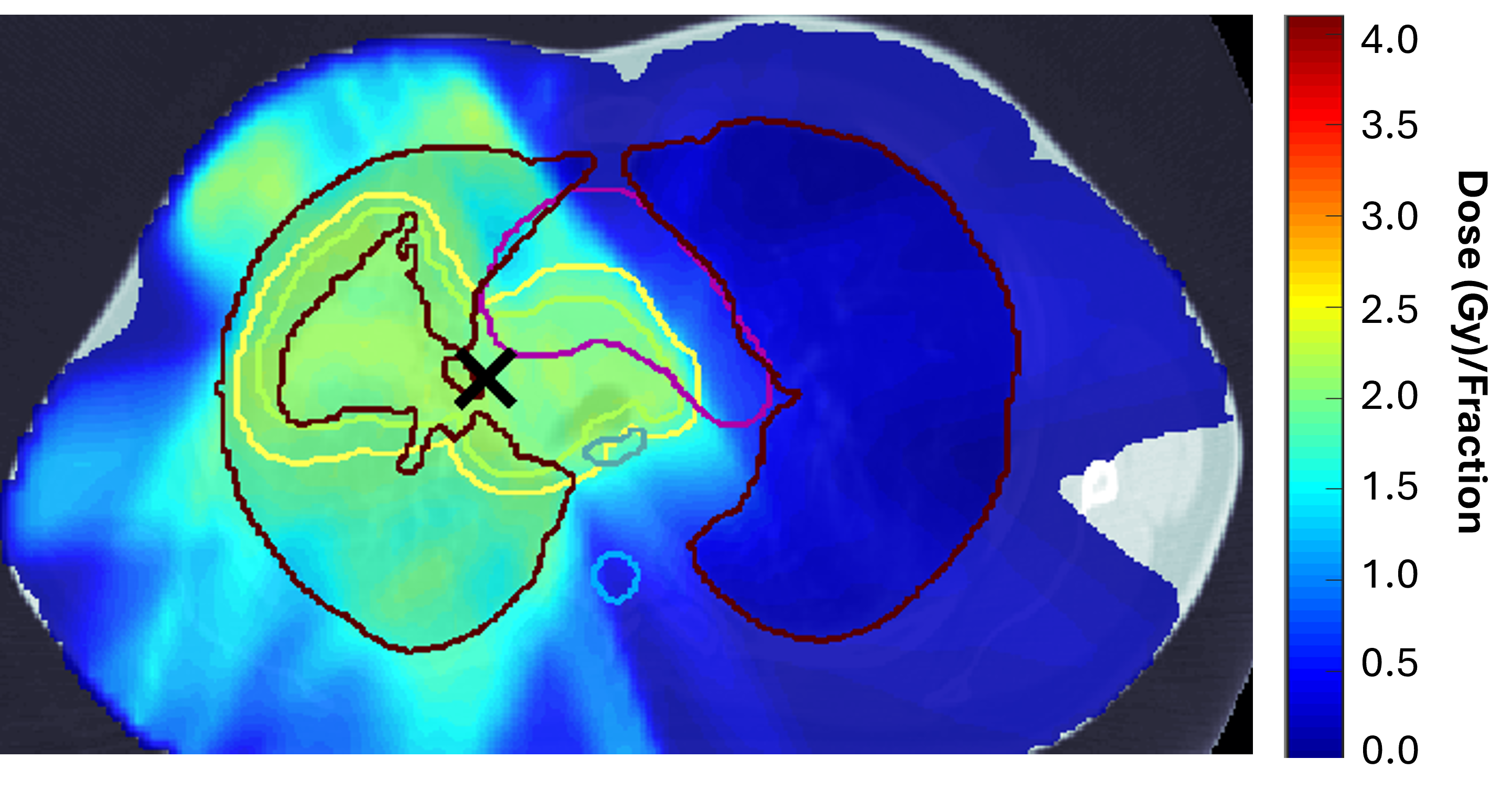}
        \caption{Risk-agnostic plan and a risk-guided plan with  only minimal trade-offs (Plan 4).}
    \end{subfigure}
    \hfill
    \begin{subfigure}[b]{0.45\textwidth}
        \centering
        \includegraphics[width=\textwidth,height=\SubFigMaxH,keepaspectratio]{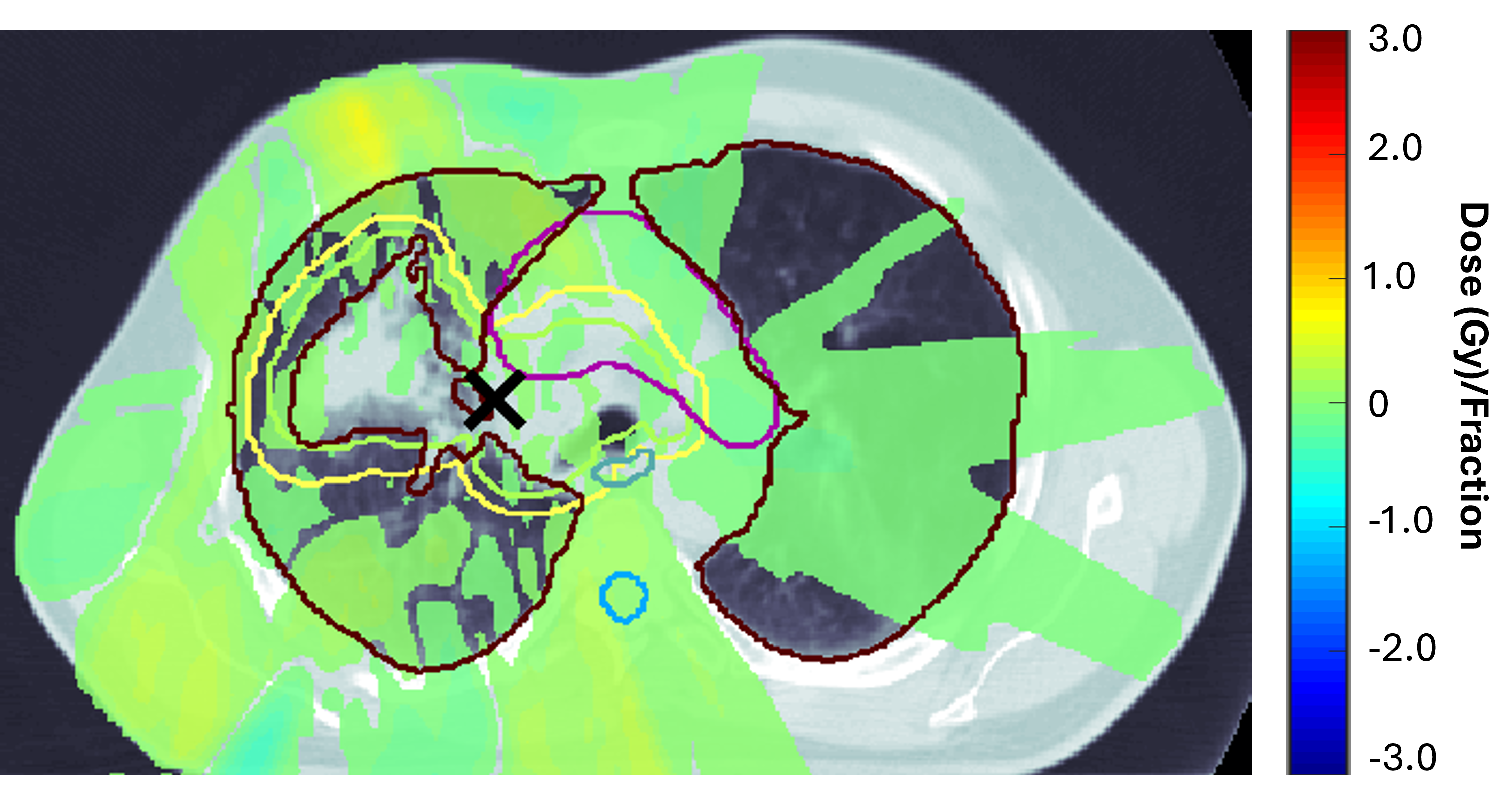}
        \caption{Difference between Plan 4 and the risk-agnostic plan}
    \end{subfigure}
    \caption{Dose distribution plots for the patient in Figure \ref{fig:difference_risk_guided_agnostic}, showing the risk-agnostic (a) and the four risk-guided plans (b, d, f, and h). The right column (c, g, e, and i) shows the dose difference (risk-guided minus the risk-agnostic plan), with negative values indicating areas of dose reduction. The contours for the CTV, PTV, total lung, heart, spinal cord, and esophagus are shown in light green, yellow, deep red, pink, blue, and grey, respectively.}
    \label{fig:difference_risk_guided_agnostic_dosewash}
\end{figure}

\subsubsection{Average risk improvement and dose trade-off across the cohort}
\label{sec:cohort_statistic}
The aggregated results of the paired comparison of a risk-agnostic and a risk-guided plan (Section \ref{sec:measuring}) in all 19 patients are shown in Table \ref{tab:statistics}. 

Over all patients, we achieved a mean risk reduction of 7.73\%, with values ranging from 0.27\% to 20.07\%.  On average, risk-guided plans show a reduction of $V5$ to the right lung of 9.48\% and a reduction of $V20$ in the total lung of 7.97\%. This is not surprising, as our model explicitly posits a positive correlation between these dose metrics and the risk. 
Lowering the lung dose to reduce the \ac{RP} risk comes at the expense of a modest increase in heart dose (mean +1.74 Gy). In all plans, the maximum spinal cord dose remains below 45 Gy. Target coverage is preserved, with \ac{CTV} D98 showing a mean difference of -1.17 Gy and –1.06 Gy at the 25th percentile.

\begin{landscape}
\begin{table}[p]
    \centering 
    \caption{Statistics for comparison of a risk-agnostic and a systematically chosen risk-guided plan per patient. Comparison is calculated as risk-guided - risk-agnostic.}
    \footnotesize
    \resizebox{\textwidth}{!}{%
    \begin{tabular}{|c|c|c|c|c|c|c|c|}
    \hline
        \makecell{Patient}
         & \makecell{Right lung \\ V5Gy [\%]} & 
         \makecell{Total lung \\ V20Gy [\%]} & 
         \makecell{Heart \\ mean [Gy]} & 
         \makecell{Spinal cord \\ V45Gy [\%]} & 
         \makecell{CTV \\ D98 [\%]} &
         \makecell{CTV \\mean [Gy]}&
         \makecell{\ac{RP} \\ risk [\%]}\\
         \hline
         1 & -5.57 & -18.30 & 0.02 & 0 & 1.64 & -0.06 &-10.39 \\
         2 & -12.45 & -5.97 &3.49 & 0 & 1.17 & -0.08 & -6.34\\
         3& -9.77 & -5.67 & 1.16 & 0 & 2.00 & -0.12 & -5.47\\
         4& -22.63 & -12.18 & 1.59 & 0 & 1.81 & -0.08 & -12.67\\
         5& -5.82 & -3.22 & 0.63 & 0 & -10.94 & 1.51 & -3.42\\ 
         6& -17.92 & -9.06 & 1.70 & 0 & 0.77 & -0.03 & -8.12\\
         7&4.26 & -7.48 & 2.42 & 0 & -2.83 & 0.22& -4.21\\
         8 & 5.22 & -9.76 &	1.15 & 0&-7.60&0.86& -5.84\\
         9 & -1.35 & 9.00 &0.28& 0	&-17.02 & 2.26 & -2.35\\
         10 &-17.55&	-15.75 & 4.97 & 0 &	0.51& -0.21 & -13.48 \\
         11 & 19.39 & -12.37 & 4.18 &0 & -1.12 & 0.23 & 	-12.55\\
         12 &-11.92 & -4.45& 0.98& 0& 2.69 & -0.04 & -6.06\\
         13 & -2.51 & -0.76 & 0.04 & 0 & 0.52 & -0.09& -0.70\\
         14 & -0.10 & -0.50& -0.06& 0 & -0.1& 0.04 & -0.27\\
         15 & -43.02& -11.91 & 4.99& 0 & -1.00 &	-0.02 & -20.07\\
         16 & -26.11& -7.65 & 0.36& 0 & 1.32& -0.17& -9.96\\
         17 & -29.10& -13.20 & 0.90& 0 & 2.91& -0.1 & -12.89\\
         18 & -3.16 & -5.90& 1.21& 0 & 2.98& 1,145 &-3.68 \\
         19 &-0.11 & -15.21& 3.03& 0 & 0.13 & 0.03 & -8.34\\ 
         \hhline{========}
         Mean diff. & -9.48	& -7.97	&1.74	&0&-1.17	&0.27	&-7.73\\
         25\% quantile &-17.74	&-12.77	&0.49	&0&-1,06	&-0.09	&-11.47\\
         75\% quantile &-0.73	&-5.06	&2,73	&0&1.72	&0.22	&-3.95\\
         std &14.35	&6.49	&1.65	&0	&5.22	&0.68	&5.09\\
        \hline
    \end{tabular}%
    }
    \label{tab:statistics}
\end{table}
\end{landscape}

\subsubsection{Contrast to conventional \ac{MCO} approach}
\label{sec:contrast_to_conventional}
In conventional \ac{MCO} as implemented in current treatment planning systems, each objective is treated equally. If, for such a method, the risk model was merely added as an equal additional objective, any dosimetric objective could be compromised to any extent (within the predefined constraints) for only marginal gains in the risk. Depending on the patient, this could easily lead to a costly calculation of a large number of plans that will not be considered by the clinician.

\begin{figure}[h]
    \centering
     \begin{subfigure}[b]{0.49\textwidth}
        \centering
        \includegraphics[width=\textwidth]{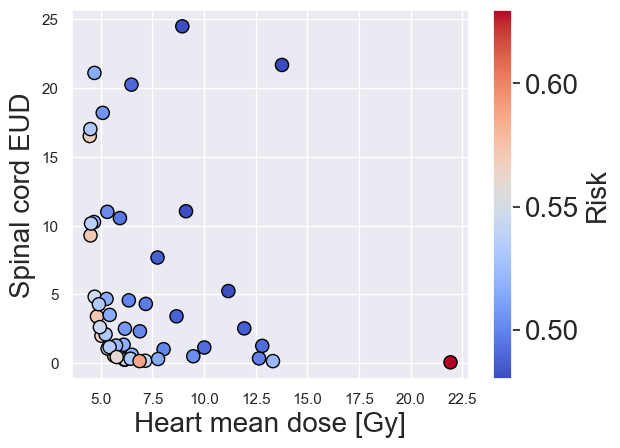}
        \caption{Conventional \ac{MCO}}
        \label{fig:pf_standard_cone}
    \end{subfigure}
    \hfill
    \begin{subfigure}[b]{0.49\textwidth}
        \centering
        \includegraphics[width=\textwidth]{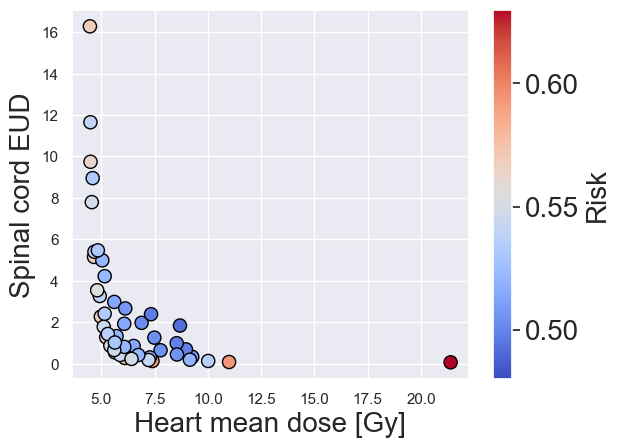}
        \caption{Our method}
        \label{fig:pf_eps10_cone}
    \end{subfigure}
    \caption{Plans created with the conventional \ac{MCO} approach, with the risk model an equal objective (left) and plans created with our approach (right). The conventional approach creates plans with high trade-offs in the dosimetric objectives. Our approach focuses plan optimization in the region close to dosimetrically optimal plans.}
    \label{fig:trad_mco_vs_our_approach}
\end{figure}

As outlined in Sections \ref{subsec:secondary_objective} and \ref{sec:adjustable_parametrization}, and illustrated in Section \ref{sec:adj_param_w_eps}, our proposed method is an improvement over the conventional \ac{MCO} methodology by allowing the planner to control the trade-off between primary (dose) and secondary (risk) objectives through the parameter $\epsilon$. Figure (\ref{fig:trad_mco_vs_our_approach}) illustrates that for a reasonably small $\epsilon$, in this case $\epsilon = 10$, our method (right) creates a different set of plans compared to conventional \ac{MCO} (left). The plans created with our method prioritize the dose objectives and allow degradations only for a considerable gain in risk. For $\epsilon \rightarrow \infty$, our method yields the same plans as conventional \ac{MCO}.

\subsubsection{Extension to three dose objectives}
\label{sec:three_objectives}

\begin{figure}
    \centering
         \begin{subfigure}[b]{0.49\textwidth}
        \centering
        \includegraphics[width=\textwidth]{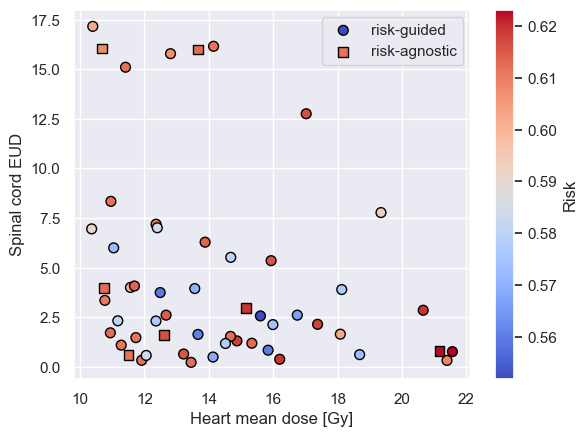}
        \caption{}
        \end{subfigure}
        \hfill
        \begin{subfigure}[b]{0.49\textwidth}
        \centering
        \includegraphics[width=\textwidth]{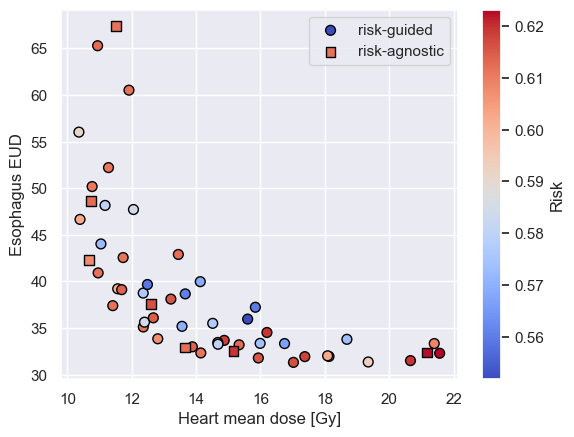}
        \caption{}
        \end{subfigure}
        \\
        \begin{subfigure}[b]{0.49\textwidth}
        \centering
        \includegraphics[width=\textwidth]{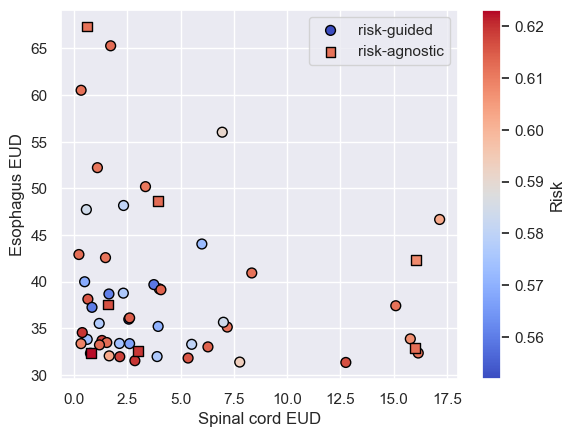}
        \caption{}
        \end{subfigure}
        \hfill
        \begin{subfigure}[b]{0.49\textwidth}
        \centering
            \includegraphics[width=\textwidth]{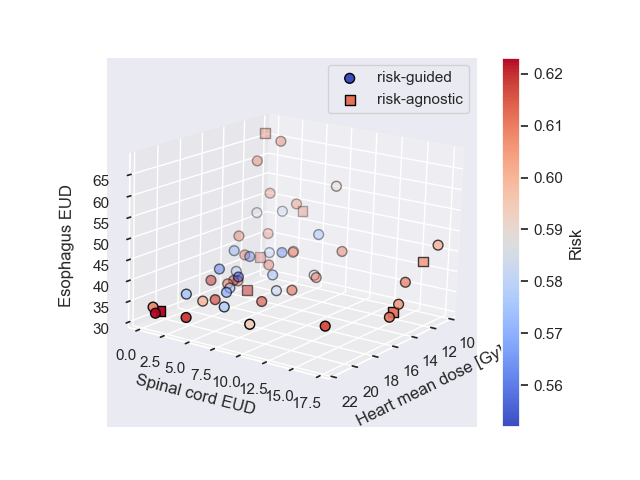}
            \caption{}
        \end{subfigure}
    \caption{Risk-guided and risk-agnostic front for three dose objectives. a)-c) two-dimensional projections of two objectives; d) three-dimensional plot of all dose objectives.}
    \label{fig:3DMCO}
\end{figure}

The previous results were based on two dose objectives. The algorithm to include the risk-optimization while maintaining a good dose distribution is compatible with arbitrary many objectives in each priority. Only the cone  $C_A$ needs to be adjusted accordingly. Adding more dose objectives not only allows the exploration of trade-offs between the different dose objectives, like in the risk-agnostic approach, it also allows to explore trade-offs between the different dose-objectives and the risk.
Figure \ref{fig:3DMCO} shows the front of a case calculated based on three objectives. Figures \ref{fig:3DMCO} a)-c) are two-dimensional projections of two out of the three objectives, d) shows a three dimensional plot of all objectives together. In addition to the mean heart dose and an \ac{EUD} function for the spinal cord, an \ac{EUD} function for the esophagus was added. As such, improvement of the predicted risk with trade-offs in all three of these objectives can be analyzed. 

For a higher number of objectives, individual plan calculations will not get significantly more expensive, as the additional evaluation of a dose objective is typically very cheap. However, even under favorable assumptions, the number of iterations needed for the Sandwiching algorithm to achieve a certain approximation quality is in the order of $\mathcal{O}(\delta^{\frac{1-N}{2}})$, with $N$ denoting the number of objectives \cite{LamKue25}. Additionally, the computational demand of the Sandwiching algorithm itself  will increase sharply with the number of objectives \cite{BokFor12}. Thus, as in conventional MCO, the number of objectives is limited by the applicability of the Sandwiching algorithm (\cite{BokFor12} mention $N = 10$ as realistic for clinical practice.).

%% file: Discussion.tex
Propelled by the advances in molecular biology, big data, and AI, the field of radiation oncology is rapidly moving towards more personalized approaches. A crucial step in this direction is moving beyond one-size-fits-all dose prescriptions which have shaped the design of RT treatment plans for decades. Patients with different comorbidity or biological profile might require unique treatment plans that reflect their needs in terms of tumor sensitivity, normal tissue tolerance, or projected immune response. This calls for direct integration of patient-specific biological profile within treatment planning paradigm. 

Our methodology is the first to directly incorporate data-driven (biological) risk models within RT treatment planning in an interactive, multi-criteria fashion, allowing for the decision maker (clinician and/or planner) to carefully weigh the pros and cons of deviating from conventional dose protocols in favor of a more individualized plan design informed by patient-specific risk of tumor outcome or treatment-induced side-effects. Acknowledging the clinical importance of established dose protocols, our approach incorporates the biological optimization as a secondary priority in plan optimization. As such, our framework can be interpreted as a re-optimization of dose-optimized plans, where risk is minimized as much as possible, subject to only minor (clinically-approved) deviations in the dose objectives.

The interactive nature of our algorithm distinguishes it from prior biologically-oriented approaches, including previous works by our team (\cite{Ajdari_2022}, \cite{Maragno_2024}), which were rigid in construction and demanded significant trust in the validity of the underlying biological risk models. Unlike prior sequential optimization approaches (\cite{Jee_2007}, \cite{Voet.2012}), our algorithm is capable of generating a family of risk-optimized plans in a \emph{single step}. This not only proved to be faster than sequential optimization \cite{Schubert.2025}, but also offers the advantage that it makes full use of the benefits of front approximation algorithms, such as the Sandwiching algorithm. Mainly, one obtains a guaranteed approximation quality that is achieved with the representative set, and interpolation between optimal points is possible while diverging at most this approximation quality from the true front. Thus, plans that were not explicitly calculated can also be obtained and considered in treatment planning.

The proposed algorithm provides the clinical decision maker with control over a key parameter ($\epsilon$) that defines acceptable relative trade-offs between the primary dose objectives and the secondary risk objective. A higher value for this parameter generates a set of alternative plans with a larger improvement in secondary criteria (risk of \ac{RP} in our example) but also increases the deviation from the conventional (dose-based) Pareto optimal plans. In this way, a higher value for $\epsilon$ allows the exploration of a wider range of trade-offs. Conversely, selecting a smaller value of $\epsilon$ imposes stricter limits on plan deviations from dose-based protocols, concentrating the resulting Pareto set on a narrower range of clinically relevant plans. As such, this parameter (indirectly) reflects and be chosen based on decision makers' trust in the underlying risk model and/or the clinical judgment about the specific case at hand. A well-validated and reliable model, or a patient deemed at a higher risk for a certain adverse outcome based on conventional planning criteria can  shift the decision towards more aggressive risk optimization (i.e., higher $\epsilon$), while less established models or borderline cases might warrant staying close to the set of conventional Pareto front (i.e., lower $\epsilon$). In this study we only used a single parameter $\epsilon$ for the relative trade-offs. However, this parameter occurs several times in the domination cone $C_A$, with the precise number depending on the dimension. It is possible to use different values here. This reflects a relative trade-off between the risk and a single objective. As such, it also allows a treatment-planner to focus on dose preservation with respect to a subset of objectives more than with respect to a different subset.

Conventional MCO can be seen as a special case of our algorithm where $\epsilon$ is infinitely large (such that any trade-off with the dose objectives is permissible). Using our method to limit $\epsilon$ narrows the span of the front, reducing the number of plan calculations needed for a good front approximation compared to the conventional approach. Limiting the range of plan alternatives also reduces the cognitive load for the planner during \ac{MCO} exploration. 

To quantify the potential risk reduction and the associated dose trade-offs, we also computed the risk-agnostic plans that exclusively consider dose objectives. In particular, plans without acceptable trade-offs in the dose objectives are inherently part of the risk-guided fronts. Thus, the additional calculation of risk-agnostic plans was performed solely for comparative purposes and is not required to identify dose-optimal plans.
In this work, both fronts were represented by a discrete set of points, a common approach in treatment planning. As a next step, navigation on an approximation of the whole front based on these representative points is often employed (see \cite{MonKue08}).  This allows treatment planners to explore the full range of trade-offs more effectively. Importantly, the use of an algorithm capable of calculating the risk-guided front in a single step facilitates this navigation.
We used a weighted sum approach to scalarize our \ac{MCO} problems, as this has proven to be an efficient way to calculate representative optimal points on the convex hull of the front. However, the risk objective and volume percentile objectives are not convex.  Thus, optimal points that are not on the convex hull might also exist. It is possible to find these points by using a different scalarization, for example the Pascoletti-Serafini scalarization \cite{PascolettiSerafini.1984}. The methodology for prioritizing dose objectives and then reoptimizing with respect to risk objectives is also compatible with other scalarizations. The arising difficulties arise only from solving a more complex, non-convex, optimization problem.  

In our risk model, current smokers showed a higher dose tolerance to \ac{RP}---a fact that is reported elsewhere (\cite{Franzen.1989}, \cite{Tucker.2008}). Consequently, the risk model predicts that a higher dose is beneficial in terms of \ac{RP} risk for these patients, and the resulting model-guided plans would recommend delivering higher lung doses to reduce the risk of RP, indicating that the model is not suitable for such patients; as such, we only included patients who are not currently smoking in our analyzes. Nevertheless, this highlights the inherent downside of purely model-driven treatment planning and emphasizes the need for balancing model-driven recommendations against established domain knowledge and clinical intuition, a fact that is baked into our proposed methodology. 

Despite its promise, our study has several limitations. First, since the main point of the present work was on developing the underling optimization algorithm, we opted for a relatively simple logistic regression as our risk model, which, compared to more sophisticated machine learning models (e.g., support vector machines, random forests, or deep learning models) might not necessarily yield the highest accuracy. However, as shown in our previous work \cite{Maragno_2024}, some of the more sophisticated ML outcome models (SVM, RF, and neural networks-based architectures) can be embedded within RT planning, which, in principle, should allow for their integration in our approach. Incorporating multiple risk models for different outcomes is also feasible. In such cases, the secondary risk optimization problem becomes an \ac{MCO} problem on its own, where slight trade-offs in the dose objectives could improve multiple risk objectives simultaneously. The degree to which each risk can be reduced depends on the compromises between competing risks. This could also be used to analyze whether an acceptable trade-off between different risks is achievable while maintaining an advantageous dose-distribution. If this is not the case, a different treatment setup or treatment modality might be indicated. The algorithm used in this study is well-suited to handle such problems with multiple secondary objectives.  Lastly, it is important to note that our study is intended as a proof-of-concept. As such, further internal and external validation is needed before the true performance and clinical benefit of our approach can be judged.

%% file: Conclusion.tex
Individual factors  influence a patient's risk for specific treatment outcomes. Incorporating these risks into treatment planning offers the potential to design plans that are better tailored to a patient's unique needs. In this work, we demonstrated how risk prediction models can be integrated into \ac{MCO} treatment planning as a secondary priority, ensuring that dose-based objectives remain the primary focus. Our findings show that such risk-guided treatment plans can achieve substantial reductions in the risk of  \ac{RP} for many \ac{NSCLC} patients, without compromising the quality of the plans with respect to dose objectives.

%% file: Appendix.tex
\subsection{Risk model for RP}
\label{sec:appendix_risk_model}

\begin{figure}
    \centering
    \includegraphics[width=\linewidth]{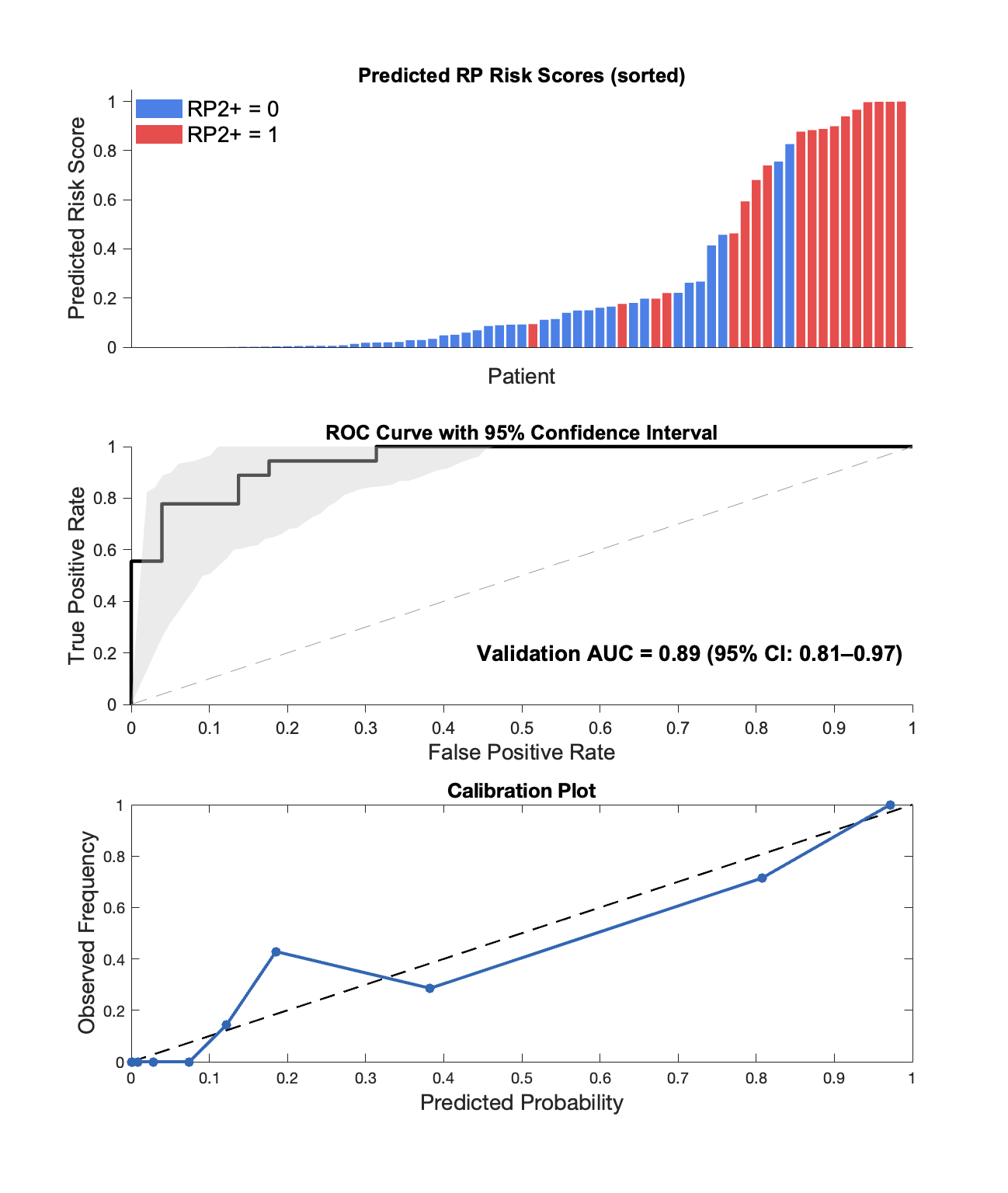}
    \caption{The results of the logistic regression model for predicting the risk of radiation pneumonitis grade 2 orhigher (RP2+).}
    \label{fig:RP_model}
\end{figure}

The logistic regression model used for predicting RP is 
\begin{equation}
    r(d) = \frac{1}{1+e^{-T(d)}},
    \label{eq:risk-model}
\end{equation}
with 
\begin{eqnarray}
T(d) &=& \ \ c_{\mathrm{PBS}}\cdot PBS + c_{\mathrm{CS}}\cdot CS + c_{\mathrm{RL}}\cdot\mathrm{V}_{\mathrm{RL}}\!\left[5Gy\right](d) + c_{\mathrm{TL}}\cdot\mathrm{V}_{\mathrm{TL}}\!\left[20Gy\right](d) \nonumber\\   
       && +c_{\mathrm{PBS,RL}}\cdot PBS\cdot\mathrm{V}_{\mathrm{RL}}\!\left[5Gy\right](d) + c_{\mathrm{CS,TL}}\cdot CS \cdot\mathrm{V}_{\mathrm{TL}}\!\left[20Gy\right](d) \nonumber\\
       && + c_{\mathrm{int}}. \nonumber
\end{eqnarray}
The coefficients of the model are recorded in Table \ref{tab:model_coefficients}. Figure \ref{fig:RP_model} shows the bootstrapped-ROC curve and the calibration plot along with the predicted risk of RP for all patients in the cohort.

\begin{table}
    \centering
    {
    \small
    \begin{tabular}{c|c|c}
         \hline
         \textbf{Type} & \textbf{Coefficient} & \textbf{Value} \\
         \hline
         intercept & $c_{\mathrm{int}}$ & -2.2937 \\
         \hline
         \multirow{4}{*}{single factor} & $c_{\mathrm{PBS}}$ & 0.5592 \\
         &  $c_{\mathrm{CS}}$ & 0.6787 \\
         &  $c_{\mathrm{RL}}$ & 3.0849 \\
         &  $c_{\mathrm{TL}}$ & 2.2056 \\
         \hline
         \multirow{2}{*}{mixed term} & $c_{\mathrm{PBS,RL}}$ & -0.8862 \\
         & $c_{\mathrm{CS,TL}}$ & -3.2135 \\
         \hline
    \end{tabular}
    } 
    \caption{List of non-zero coefficients for the logistic regression model \ref{eq:risk-model} with factors for pre-treatment breathing status ($PBS$), current smoking status ($CS$), right lung V5Gy ($RL$), and total lung V20Gy ($TL$).}
    \label{tab:model_coefficients}
\end{table}

%% file: References.bib
@article{Ajdari_2022,
doi = {10.1088/1361-6560/ac88b3},
url = {https://doi.org/10.1088/1361-6560/ac88b3},
year = {2022},
month = {sep},
publisher = {IOP Publishing},
volume = {67},
number = {18},
pages = {185015},
author = {Ajdari, Ali and Liao, Zhongxing and Mohan, Radhe and Wei, Xiong and Bortfeld, Thomas},
title = {Personalized mid-course FDG-PET based adaptive treatment planning for non-small cell lung cancer using machine learning and optimization},
journal = {Physics in Medicine \& Biology}
}

@article{Bera.2022,
 author = {Bera, Kaustav and Braman, Nathaniel and Gupta, Amit and Velcheti, Vamsidhar and Madabhushi, Anant},
 year = {2022},
 title = {Predicting cancer outcomes with radiomics and artificial intelligence in radiology},
 pages = {132--146},
 volume = {19},
 number = {2},
 issn = {1759-4782},
 journal = {Nature Reviews Clinical Oncology},
 doi = {10.1038/s41571-021-00560-7}
}

@article{BokFor12,
  title={An Algorithm for Approximating Convex Pareto Surfaces Based on Dual Techniques},
  author={Bokrantz, Rasmus and Forsgren, Anders},
  journal={Informs Journal on Computing},
  volume={25},
  number = {2},
  year={2012}
}

@article{BreStoVoe12,
    author = {Breedveld,~S. and Storchi,~P.~R. and Voet,~P.~W. and Heijmen,~B.~J.},
    title = {iCycle: Integrated, multicriterial beam angle, and profile optimization for generation of coplanar and noncoplanar IMRT plans.},
    journal = {Medical physics},
    volume = {39(2)},
    year = {2012}
}

@Inbook{ByrNocWal06,
author={Byrd, Richard H. and Nocedal, Jorge and Waltz, Richard A.},
title={Knitro: An Integrated Package for Nonlinear Optimization},
bookTitle={Large-Scale Nonlinear Optimization},
year={2006},
publisher={Springer US},
pages={35--59}
}

@Inbook{CisMaiZie15,
  author={Cisternas,~E. and Mairani,~A. and Ziegenhein,~P. and J{\"{a}}kel,~O. and Bangert,~M.},
  title={matRad - a multi-modality open source 3D treatment planning toolkit},
  bookTitle={World Congress on Medical Physics and Biomedical Engineering, June 7-12, 2015, Toronto, Canada}, 
  year={2015},
  pages={1608--1611}, 
  publisher={Springer International Publishing}
}

@article{Deist.2018,
author = {Deist, Timo M. and Dankers, Frank J. W. M. and Valdes, Gilmer and Wijsman, Robin and Hsu, I-Chow and Oberije, Cary and Lustberg, Tim and van Soest, Johan and Hoebers, Frank and Jochems, Arthur and El Naqa, Issam and Wee, Leonard and Morin, Olivier and Raleigh, David R. and Bots, Wouter and Kaanders, Johannes H. and Belderbos, José and Kwint, Margriet and Solberg, Timothy and Monshouwer, René and Bussink, Johan and Dekker, Andre and Lambin, Philippe},
title = {Machine learning algorithms for outcome prediction in (chemo)radiotherapy: An empirical comparison of classifiers},
journal = {Medical Physics},
volume = {45},
number = {7},
pages = {3449-3459},
doi = {https://doi.org/10.1002/mp.12967},
url = {https://aapm.onlinelibrary.wiley.com/doi/abs/10.1002/mp.12967},
eprint = {https://aapm.onlinelibrary.wiley.com/doi/pdf/10.1002/mp.12967},
year = {2018}
}

@BOOK{eaton:2002,
    author =     "John W. Eaton",
    title =      "GNU Octave Manual",
    publisher =  "Network Theory Limited",
    year =       "2002",
    isbn =       "0-9541617-2-6"
}

@book{Ehr05,
	author = {Ehrgott, Matthias},
	year = {2005},
	title = {Multicriteria Optimization},
	url = {https://ebookcentral.proquest.com/lib/kxp/detail.action?docID=6311706},
	address = {Berlin, Heidelberg},
	edition = {Second edition},
	publisher = {{Springer Berlin · Heidelberg}},
	isbn = {9783540276593}
}

@article{Eic09,
  author = {Eichfelder, Gabriele},
  title = {Scalarizations for adaptively solving multi-objective optimization problems},
  journal = {Compututational Optimization and Applications},
  year = {2009},
  volume = {44},
  pages = {249--273}
}

@article{Franzen.1989,
author = {Lars Franz\'{e}n and Leif Bjermer and Roger Henriksson and Bo Littbrand and Kenneth Nilsson},
title = {Does Smoking Protect Against Radiation-induced Pneumonitis?},
journal = {International Journal of Radiation Biology},
volume = {56},
number = {5},
pages = {721--724},
year = {1989},
publisher = {Taylor \& Francis},
doi = {10.1080/09553008914551961},
note ={PMID: 2573669},
URL = {https://doi.org/10.1080/09553008914551961},
eprint = {https://doi.org/10.1080/09553008914551961}}

@article{Jee_2007,
doi = {10.1088/0031-9155/52/7/006},
url = {https://doi.org/10.1088/0031-9155/52/7/006},
year = {2007},
month = {mar},
publisher = {},
volume = {52},
number = {7},
pages = {1845},
author = {Jee, Kyung-Wook and McShan, Daniel L and Fraass, Benedick A},
title = {Lexicographic ordering: intuitive multicriteria optimization for IMRT},
journal = {Physics in Medicine \& Biology}}

@article{Kierkels.2016,
title = {Multivariable normal tissue complication probability model-based treatment plan optimization for grade 2–4 dysphagia and tube feeding dependence in head and neck radiotherapy},
journal = {Radiotherapy and Oncology},
volume = {121},
number = {3},
pages = {374-380},
year = {2016},
issn = {0167-8140},
doi = {https://doi.org/10.1016/j.radonc.2016.08.016},
url = {https://www.sciencedirect.com/science/article/pii/S0167814016342803},
author = {Roel G.J. Kierkels and Kim Wopken and Ruurd Visser and Erik W. Korevaar and Arjen {van der Schaaf} and Hendrik P. Bijl and Johannes A. Langendijk},
}

@article{LamKue25,
    author = {Lammel, I. and Küfer, K. H. and Süss, P.},
    title = {Efficient Approximation Quality Computation for Sandwiching Algorithms for Convex Multicriteria Optimization.},
    journal = {Journal of Optimization Theory and Applications},
    year = {2025},
    volume = {204(3)}
}

@article{Li.2024,
 author = {Li, Shen and He, Yuxin and Liu, Jifeng and Chen, Kefan and Yang, Yuzhao and Tao, Kai and Yang, Jiaqing and Luo, Kui and Ma, Xuelei},
 year = {2024},
 title = {An umbrella review of socioeconomic status and cancer},
 pages = {9993},
 volume = {15},
 number = {1},
 issn = {2041-1723},
 journal = {Nature Communications},
 doi = {10.1038/s41467-024-54444-2}
}

@article{Liao.2018,
author = {Liao, Zhongxing and Lee, J. Jack and Komaki, Ritsuko and Gomez, Daniel R. and O’Reilly, Michael S. and Fossella, Frank V. and Blumenschein, George R. and Heymach, John V. and Vaporciyan, Ara A. and Swisher, Stephen G. and Allen, Pamela K. and Choi, Noah Chan and DeLaney, Thomas F. and Hahn, Stephen M. and Cox, James D. and Lu, Charles S. and Mohan, Radhe },
title = {Bayesian Adaptive Randomization Trial of Passive Scattering Proton Therapy and Intensity-Modulated Photon Radiotherapy for Locally Advanced Non–Small-Cell Lung Cancer},
journal = {Journal of Clinical Oncology},
volume = {36},
number = {18},
pages = {1813-1822},
year = {2018},
doi = {10.1200/JCO.2017.74.0720},
note ={PMID: 29293386},
URL = {https://ascopubs.org/doi/abs/10.1200/JCO.2017.74.0720},
eprint = {https://ascopubs.org/doi/pdf/10.1200/JCO.2017.74.0720}}

@article{Maragno_2024,
doi = {10.1088/1361-6560/ad2d7e},
url = {https://doi.org/10.1088/1361-6560/ad2d7e},
year = {2024},
month = {mar},
publisher = {IOP Publishing},
volume = {69},
number = {7},
pages = {075003},
author = {Maragno, Donato and Buti, Gregory and Birbil, {\c{S}}.  {\.{I}}lker and Liao, Zhongxing and Bortfeld, Thomas and den Hertog, Dick and Ajdari, Ali},
title = {Embedding machine learning based toxicity models within radiotherapy treatment plan optimization},
journal = {Physics in Medicine \& Biology}
}

@article{MonKue08,
  author = {Monz,~M. and K{\"{u}}fer,~K.~H. and Bortfeld,~T.~R. and Thieke,~C.},
  title = {Pareto navigation - algorithmic foundation of interactive multi-criteria {IMRT} planning},
  journal = {Phys. Med. Biol.},
  year = {2008},
  volume = {53},
  pages = {985-998}
}

@article{Pan.2023,
 author = {Pan, Yi and Zhang, Jia-Tao and Gao, Xuan and Chen, Zhi-Yong and Yan, Bingfa and Tan, Pei-Xin and Yang, Xiao-Rong and Gao, Wei and Gong, Yuhua and Tian, Zihan and Liu, Si-Yang Maggie and Lin, Hui and Sun, Hao and Huang, Jie and Liu, Si-Yang and Yan, Hong-Hong and Dong, Song and Xu, Chong-Rui and Chen, Hua-Jun and Wang, Zhen and Li, Pansong and Guan, Yanfang and Wang, Bin-Chao and Yang, Jin-Ji and Tu, Hai-Yan and Yang, Xue-Ning and Zhong, Wen-Zhao and Xia, Xuefeng and Yi, Xin and Zhou, Qing and Wu, Yi-Long},
 year = {2023},
 title = {Dynamic circulating tumor DNA during chemoradiotherapy predicts clinical outcomes for locally advanced non-small cell lung cancer patients},
 pages = {1763-1773.e4},
 volume = {41},
 number = {10},
 issn = {1535-6108},
 journal = {Cancer Cell},
 doi = {10.1016/j.ccell.2023.09.007}
}

@article{PascolettiSerafini.1984,
 author = {Pascoletti, A. and Serafini, P.},
 year = {1984},
 title = {Scalarizing Vector Optimization Problems},
 pages = {499--524},
 volume = {42},
 number = {4},
 journal = {Journal of Optimization Theory and Applications},
 doi = {10.1007/BF00934564}
}

@article{Peeken.2017,
 author = {Peeken, Jan Caspar and N\"{u}sslin, Fridtjof and Combs, Stephanie E.},
 year = {2017},
 title = {“Radio-oncomics”},
 pages = {767--779},
 volume = {193},
 number = {10},
 issn = {1439-099X},
 journal = {Strahlentherapie und Onkologie},
 doi = {10.1007/s00066-017-1175-0}
}

@article{Petkar.2025,
title = {De-escalation of adjuvant radiotherapy in HPV-associated oropharyngeal cancer},
journal = {The Lancet Oncology},
volume = {26},
number = {9},
pages = {1127-1129},
year = {2025},
issn = {1470-2045},
doi = {https://doi.org/10.1016/S1470-2045(25)00408-5},
url = {https://www.sciencedirect.com/science/article/pii/S1470204525004085},
author = {Imran Petkar}
}

@techreport{RayMCO17,
    author = {RaySearch},
    title = {Multi-criteria optimization for tomotherapy},
    institution = {RaySearch},
    year = {2017}
}

@article{Rabasco.2022,
 author = {{Rabasco Meneghetti}, Asier and Zwanenburg, Alex and Linge, Annett and Lohaus, Fabian and Grosser, Marianne and Baretton, Gustavo B. and Kalinauskaite, Goda and Tinhofer, Ingeborg and Guberina, Maja and Stuschke, Martin and Balermpas, Panagiotis and von der Grün, Jens and Ganswindt, Ute and Belka, Claus and Peeken, Jan C. and Combs, Stephanie E. and Böke, Simon and Zips, Daniel and Troost, Esther G. C. and Krause, Mechthild and Baumann, Michael and Löck, Steffen},
 year = {2022},
 title = {Integrated radiogenomics analyses allow for subtype classification and improved outcome prognosis of patients with locally advanced HNSCC},
 pages = {16755},
 volume = {12},
 number = {1},
 journal = {Scientific reports},
 doi = {10.1038/s41598-022-21159-7}
}

@article{Raturi.2021,
    author = {Raturi, Vijay P and Motegi, Atsushi and Zenda, Sadamoto and Nakamura, Naoki and Hojo, Hidehiro and Kageyama, Shin-Ichiro and Okumura, Masayuki and Rachi, Toshiya and Ohyoshi, Hajime and Tachibana, Hidenobu and Motegi, Kana and Ariji, Takaki and Nakamura, Masaki and Hirano, Yasuhiro and Hirata, Hidenari and Akimoto, Tetsuo},
    title = {Comparison of a Hybrid IMRT/VMAT technique with non-coplanar VMAT and non-coplanar IMRT for unresectable olfactory neuroblastoma using the RayStation treatment planning system—EUD, NTCP and planning study},
    journal = {Journal of Radiation Research},
    volume = {62},
    number = {3},
    pages = {540-548},
    year = {2021},
    month = {04},
    issn = {1349-9157},
    doi = {10.1093/jrr/rrab010},
    url = {https://doi.org/10.1093/jrr/rrab010},
    eprint = {https://academic.oup.com/jrr/article-pdf/62/3/540/37907407/rrab010.pdf},
}

@article{Ren.2023,
author = {Ren, Kang and Shen, Lin and Qiu, Jianfeng and Sun, Kui and Chen, Tingyin and Xuan, Long and Yang, Minwu and She, Hao-Yuan and Shen, Liangfang and Zhu, Hong and Deng, Lan and Jing, Di and Shi, Liting},
title = {Treatment planning computed tomography radiomics for predicting treatment outcomes and haematological toxicities in locally advanced cervical cancer treated with radiotherapy: A retrospective cohort study},
journal = {BJOG: An International Journal of Obstetrics \& Gynaecology},
volume = {130},
number = {2},
pages = {222-230},
doi = {https://doi.org/10.1111/1471-0528.17285},
url = {https://obgyn.onlinelibrary.wiley.com/doi/abs/10.1111/1471-0528.17285},
eprint = {https://obgyn.onlinelibrary.wiley.com/doi/pdf/10.1111/1471-0528.17285},
year = {2023}}

@article{Rydzewski.2023,
title = {Machine Learning \& Molecular Radiation Tumor Biomarkers},
journal = {Seminars in Radiation Oncology},
volume = {33},
number = {3},
pages = {243-251},
year = {2023},
note = {Dedicated Issue on Personalized Dose in Radiation Oncology},
issn = {1053-4296},
doi = {https://doi.org/10.1016/j.semradonc.2023.03.002},
url = {https://www.sciencedirect.com/science/article/pii/S1053429623000164},
author = {Nicholas R. Rydzewski and Kyle T. Helzer and Matthew Bootsma and Yue Shi and Hamza Bakhtiar and Martin Sjöström and Shuang G. Zhao}}

@article{Schubert.2025,
 author = {Schubert, Mara and Teichert, Katrin},
 year = {2025},
 title = {Bi‐Level Multi‐Criteria Optimization to Include Linear Energy Transfer Into Proton Treatment Planning},
 volume = {32},
 number = {3},
 issn = {1057-9214},
 journal = {Journal of Multi-Criteria Decision Analysis},
 doi = {10.1002/mcda.70021}
}

@phdthesis{Ser12,
  author={J I~Serna},
  title={Multi-objective Optimization in Mixed Integer Problems with Application to the Beam Angle Optimization Problem in IMRT},
  school={Technical University of Kaiserslautern, Department of Mathematics},
  year={2012}
}

@article{Toma-Dasu.2013,
author = {Toma-Dasu, Iuliana and Dasu, Alexandru},
title = {Modelling Tumour Oxygenation, Reoxygenation and Implications on Treatment Outcome},
journal = {Computational and Mathematical Methods in Medicine},
volume = {2013},
number = {1},
pages = {141087},
doi = {https://doi.org/10.1155/2013/141087},
url = {https://onlinelibrary.wiley.com/doi/abs/10.1155/2013/141087},
eprint = {https://onlinelibrary.wiley.com/doi/pdf/10.1155/2013/141087},
year = {2013}
}

@article{Tucker.2008,
 author = {Tucker, Susan L. and Liu, H. Helen and Liao, Zhongxing and Wei, Xiong and Wang, Shulian and Jin, Hekun and Komaki, Ritsuko and Martel, Mary K. and Mohan, Radhe},
 year = {2008},
 title = {Analysis of Radiation Pneumonitis Risk Using a Generalized Lyman Model},
 pages = {568--574},
 volume = {72},
 number = {2},
 journal = {International journal of radiation oncology, biology, physics},
 doi = {10.1016/j.ijrobp.2008.04.053}
}

@InProceedings{Ureba.2022,
author="Ureba, Ana
and {\"O}d{\'e}n, Jakob
and Toma-Dasu, Iuliana
and Lazzeroni, Marta",
editor="Scholkmann, Felix
and LaManna, Joseph
and Wolf, Ursula",
title="Photon and Proton Dose Painting Based on Oxygen Distribution -- Feasibility Study and Tumour Control Probability Assessment",
booktitle="Oxygen Transport to Tissue XLIII",
year="2022",
publisher="Springer International Publishing",
address="Cham",
pages="223--228",
isbn="978-3-031-14190-4"
}

@techreport{VarMCO19,
    author = {Varian},
    title = {Multi-Criteria Optimization: Creating High-Quality Treatment Plans in a Fraction of the Time},
    institution = {https://www.varian.com/multi-criteria-optimization-creating-high-quality-treatment},
    year = {2019}
}

@article{Velu.2024,
 author = {Velu, Umesh and Singh, Anshul and Nittala, Roselin and Yang, Johnny and Vijayakumar, Srinivasan and Cherukuri, Chanukya and Vance, Gregory R. and Salvemini, John D. and Hathaway, Bradley F. and Grady, Camille and Roux, Jeffrey A. and Lewis, Shirley},
 year = {2024},
 title = {Precision Population Cancer Medicine in Brain Tumors: A Potential Roadmap to Improve Outcomes and Strategize the Steps to Bring Interdisciplinary Interventions},
 pages = {e71305},
 volume = {16},
 number = {10},
 issn = {2168-8184},
 journal = {Cureus},
 doi = {10.7759/cureus.71305}
}

@article{Voet.2012,
author = {Voet, Peter W. J. and Breedveld, Sebastiaan and Dirkx, Maarten L. P. and Levendag, Peter C. and Heijmen, Ben J. M.},
title = {Integrated multicriterial optimization of beam angles and intensity profiles for coplanar and noncoplanar head and neck IMRT and implications for VMAT},
journal = {Medical Physics},
volume = {39},
number = {8},
pages = {4858-4865},
year={2012},
doi = {https://doi.org/10.1118/1.4736803},
url = {https://aapm.onlinelibrary.wiley.com/doi/abs/10.1118/1.4736803},
eprint = {https://aapm.onlinelibrary.wiley.com/doi/pdf/10.1118/1.4736803}}

@article{Wang.25,
author = {Wang, Wei and Li, Wangyao and Li, Jiaxin and Lin, Yuting and Liu, Xu and Qin, Bin and Gao, Hao},
title = {Direct minimization of normal-tissue toxicity via an NTCP-based IMPT planning method},
journal = {Medical Physics},
volume = {52},
number = {3},
pages = {1399-1407},
doi = {https://doi.org/10.1002/mp.17559},
url = {https://aapm.onlinelibrary.wiley.com/doi/abs/10.1002/mp.17559},
eprint = {https://aapm.onlinelibrary.wiley.com/doi/pdf/10.1002/mp.17559},
year = {2025}
}

@article{Wei.2019,
 author = {Wei, Lise and Osman, Sarah and Hatt, Mathieu and {El Naqa}, Issam},
 year = {2019},
 title = {Machine learning for radiomics-based multimodality and multiparametric modeling},
 pages = {323--338},
 volume = {63},
 number = {4},
 issn = {1824-4785},
 journal = {The quarterly journal of nuclear medicine and molecular imaging},
 doi = {10.23736/S1824-4785.19.03213-8}
}

@article{Xiao.2025,
author = {Lihong Xiao and Youhua Wang and Xiangxiang Shi and Haowen Pang and Yunfei Li},
title ={Computed tomography–based radiomics modeling to predict patient overall survival in cervical cancer with intensity-modulated radiotherapy combined with concurrent chemotherapy},
journal = {Journal of International Medical Research},
volume = {53},
number = {3},
pages = {03000605251325996},
year = {2025},
doi = {10.1177/03000605251325996},
note ={PMID: 40119689},
URL = {https://doi.org/10.1177/03000605251325996},
eprint = {https://doi.org/10.1177/03000605251325996}}

@article{Yuan.2025,
author = {Yuan, Xiaohan and Ma, Chaoqiong and Hu, Mingzhe and Qiu, Richard L. J. and Salari, Elahheh and Martini, Reema and Yang, Xiaofeng},
title = {Machine learning in image-based outcome prediction after radiotherapy: A review},
journal = {Journal of Applied Clinical Medical Physics},
volume = {26},
number = {1},
pages = {e14559},
doi = {https://doi.org/10.1002/acm2.14559},
url = {https://aapm.onlinelibrary.wiley.com/doi/abs/10.1002/acm2.14559},
eprint = {https://aapm.onlinelibrary.wiley.com/doi/pdf/10.1002/acm2.14559},
year = {2025}
}

@software{MATLAB,
    year = {2022},
    author = {The MathWorks Inc.},
    title = {MATLAB version: 9.13.0 (R2022b)},
    publisher = {The MathWorks Inc.},
    address = {Natick, Massachusetts, United States},
    url = {https://www.mathworks.com}
}
